\documentclass[a4paper,12pt]{article}
\usepackage{latexsym,amsmath,amsfonts,amssymb}
\usepackage[latin1]{inputenc}
\usepackage[american]{babel}
\usepackage[dvips]{graphicx}
\usepackage{bbm}

\newcommand{\re}{\,\mathbb{R}\mbox{e}\,}
\newcommand{\im}{\,\mathbb{I}\mbox{m}\,}

\newcommand{\dvol}{d\mathrm{vol}}

\newcommand{\Nugual}[1]{$\mathcal{N}= #1 $}

\newcommand{\parfrac}[2]{\frac{\partial #1}{\partial #2}}

\newcommand{\ie}{\textit{i.e.} }

\numberwithin{equation}{section}

\newcommand{\be}{\begin{equation}} \newcommand{\ee}{\end{equation}}
\newcommand{\bea}{\begin{equation} \begin{aligned}} \newcommand{\eea}{\end{aligned} \end{equation}}

\newcommand{\calC}{\mathcal{C}}
\newcommand{\calD}{\mathcal{D}}

\newcommand{\calN}{\mathcal{N}}
\newcommand{\calO}{\mathcal{O}}
\newcommand{\calQ}{\mathcal{Q}}

\newcommand{\bbC}{\mathbb{C}}

\newcommand{\bbZ}{\mathbb{Z}}

\begin{document}

\makeatletter \@addtoreset{equation}{section} \makeatother
\renewcommand{\theequation}{\thesection.\arabic{equation}}
\pagestyle{empty}
\rightline{PUPT-2285}
\rightline{SISSA-68/2008/EP}
\rightline{TAUP-2886/08}
\vspace{0.8cm}
\begin{center}
{\LARGE{\bf The $\mathbf{{\cal N}=2}$ cascade revisited \vskip 7pt and
the enhan\c{c}on bearings \\[10mm]}} {\large{Francesco Benini$^{1,2}$,
Matteo Bertolini$^{2}$, Cyril Closset$^{3}$ \\ and  Stefano Cremonesi$^{2,4}$ \\[5mm]}}
{\small{{}$^1$ Department of Physics, Princeton University \\
\vspace*{-2pt} Princeton, NJ 08544, USA\\

\medskip
{}$^2$ SISSA and INFN - Sezione di Trieste\\ 
\vspace*{-2pt} Via Beirut 2; I 34014 Trieste, Italy\\

\medskip
{}$^3$ Physique Th\'eorique et Math\'ematique and International Solvay
Institutes \\ 
\vspace*{-2pt} Universit\'e Libre de Bruxelles, C.P. 231, 1050
Bruxelles, Belgium \\

\medskip
{}$^4$
Raymond and Beverly Sackler Faculty of Exact Sciences, School of Physics and Astronomy\\
\vspace*{-2pt} Tel-Aviv University, Ramat-Aviv 69978, Israel}}

\medskip

\medskip

\medskip

\medskip

\medskip

{\bf Abstract}
\vskip 20pt
\begin{minipage}[h]{16.0cm}

Supergravity backgrounds with varying fluxes generated by fractional
branes at non-isolated Calabi-Yau singularities had escaped a precise
dual field theory interpretation so far. In the present work, considering
the prototypical example of such models, the $\mathbb{C} \times
\mathbb{C}^2/\bbZ_2$ orbifold, we propose a solution for this problem,
and show that the known cascading solution corresponds to a vacuum on the
Coulomb branch of the corresponding quiver gauge theory involving a
sequence of strong coupling transitions reminiscent of the baryonic root
of $\calN=2$ SQCD.  We also find a slight modification of this cascading vacuum 
which upon mass deformation is expected to flow to the Klebanov-Strassler 
cascade. Finally, we discuss an infinite class of vacua on the
Coulomb branch whose RG flows include infinitely coupled conformal
regimes, and explain their gravitational manifestation in terms of new
geometric structures that we dub enhan\c con bearings. Repulson-free
backgrounds dual to all the vacua we analyse are explicitly provided.

\end{minipage}
\end{center}
\newpage
\setcounter{page}{1} \pagestyle{plain}
\renewcommand{\thefootnote}{\arabic{footnote}} \setcounter{footnote}{0}

\tableofcontents

\vspace*{1cm}

\section{Introduction and summary}
\label{intro}

Supergravity solutions with running fluxes are ubiquitous in non-conformal versions 
of the gauge/gravity correspondence. In fact, they occur whenever fractional branes 
are present. The decrease of such fluxes as a function of the holographic coordinate 
is believed to correspond to a reduction in the number of degrees of freedom of the 
dual gauge theory as the latter flows towards the IR.

The most widely known example is the famous
Klebanov-Tseytlin-Strassler model, arising from fractional branes at a
conifold singularity \cite{Gubser:1998fp, Klebanov:1999rd,
Klebanov:2000nc,Klebanov:2000hb}. In such a context the dual gauge
theory interpretation of running fluxes is in terms of a
renormalization group (RG) flow described by a cascade of Seiberg
dualities occurring at subsequent strong coupling scales and lowering
the rank of the strongly coupled gauge group (see
\cite{Strassler:2005qs} for a review). This flow takes place along the baryonic branch of the gauge theory. 
The low energy dynamics involves confinement and chiral symmetry breaking, which geometrically translate into a 
complex structure deformation of the
singularity. This behavior is prototypical of any isolated singularity
admitting complex structure deformation.

By placing fractional branes at isolated singularities with obstructed complex structure 
deformation \cite{Berenstein:2005xa, Franco:2005zu, Bertolini:2005di} one obtains theories 
whose RG flow is expected to be similarly described by a cascade of Seiberg dualities, but where 
the geometric obstruction translates into a runaway along a baryonic direction \cite{Intriligator:2005aw}.

The case of fractional branes at non-isolated singularities, which involves twisted sector 
fields propagating along the complex line singularity, was less understood so far. 
The simplest such example, which is a \Nugual{2} model 
obtained considering fractional branes at a $\bbC \times \bbC^2/\bbZ_2$ orbifold (also
known as $A_1$ singularity) \cite{Bertolini:2000dk, Polchinski:2000mx}, 
has been interpreted in various ways in the literature
\cite{Polchinski:2000mx, Aharony:2000pp,
Petrini:2001fk}. Consideration of probe fractional branes in the
supergravity solutions \cite{Polchinski:2000mx} and recent methods
based on the computation of Page charges \cite{Benini:2007gx,Argurio:2008mt} 
suggest that the RG flow of the dual theories involves 
strong coupling transitions where the rank of the non-abelian factor
in a gauge group with an adjoint chiral superfield drops according to
the same numerology as in Seiberg duality, leading to a cascade. Since
Seiberg-like dualities do not hold in this case, such strong coupling
transitions cry for an explanation. It is worth stressing that such a
phenomenon is not specific to $\calN=2$ models, but instead appears quite generically in any 
\Nugual{1} setup admitting non-isolated singularities together with isolated 
ones: the RG flow, as read from the gravity solution, is described by suitable 
combinations of Seiberg duality cascades and \Nugual{2}-like transitions
\cite{Argurio:2008mt}. Therefore, clarifying which field theory dynamics governs
these transitions is instrumental to understanding how string theory UV-completes 
field theories arising on systems of fractional branes at rather 
generic CY singularities. 

To that aim, in this paper we reconsider the cascading solution describing regular and 
fractional D3 branes at the $\bbC \times \bbC^2/\bbZ_2$ orbifold, as a prototype of the 
more general class of branes at non-isolated singularities, and provide a solution for 
this problem. Our proposal elaborates on previous ones \cite{Polchinski:2000mx,Aharony:2000pp}, 
and  solves a number of problems raised there. The dual gauge
theory is a $SU(N+M)\times
SU(N)$ $\calN=2$ quiver with bifundamental matter, where $N$ is the
number of regular branes and $M$ the number of fractional ones, and
its dual supergravity solution is known \cite{Bertolini:2000dk}. The
structure of such a gauge theory has many similarities with the
conifold one, and the two are indeed related by a $\calN=1$-preserving
mass deformation \cite{Klebanov:1998hh}. In order to provide a precise
interpretation of the cascading RG flow, we start approximating the
dynamics around scales where one of the two gauge coupling diverges
with an effective \Nugual{2} SQCD, treating the other group as
global. This allows us to claim that the transition occurs at the
baryonic root (\ie the point of the quantum moduli space of $\calN=2$
SQCD where the baryonic branch meets the Coulomb branch), where the
strongly coupled $SU(N+M)$ group is effectively broken to $SU(N-M)$
(plus abelian factors). As in the $\calN=1$ conifold model, this is an
iterative process which has the effect of lowering the effective
ranks of the two gauge groups as the energy decreases, in a way
which is exactly matched by the dual supergravity solution. On the
other hand, the power of the Seiberg-Witten (SW) curve technology allows us
to check our claim exactly, in the full quiver theory.

Models arising from branes at non-isolated singularities have the
distinctive property of having, besides a Higgs branch, also a Coulomb
branch. This allows for a rather mundane UV completion of the
cascading quiver theory, starting with the conformal $SU(N+M) \times
SU(N+M)$ theory engineered by $N+M$ D3 branes at the orbifold
singularity, and Higgsing it at some scale $z_0$
\cite{Polchinski:2000mx}. This stops the cascade in the UV as the theory 
is in a  superconformal phase at energies higher than $z_0$ 
(notice that such a simple SCFT completion is not possible for the 
$\calN=1$ conifold model; see \cite{Hollowood:2004ek} for alternative ways to 
UV-complete the $\calN=1$ 
cascade with a SCFT). We first
discuss the case where the cutoff is at finite energy: by means of
the relevant Seiberg-Witten curves \cite{Seiberg:1994aj,
Seiberg:1994rs}, we provide a detailed analysis of several vacua on
the Coulomb branch, together with the corresponding supergravity
duals. For vacua at the origin of the Coulomb branch, there is in fact
no cascade at all \cite{Petrini:2001fk}, while we show that the
smaller is the number of adjoints fields having vanishing VEV, the
larger is the number of steps in the cascade.

We then consider the case where the cutoff is sent to infinity,
corresponding to the infinite cascade limit. This setup is the one
which makes contact with the conifold cascade, as the two are expected
to be related by a mass deformation. Actually, only specific vacua of
the $\calN=2$ theory survive such a mass deformation
\cite{Argyres:1996eh}, and we provide the corresponding SW curve, with
a parametrically high level of accuracy. To find the supergravity
solution interpolating from the $\calN=2$ to the $\calN=1$ cascade is
left to future research.

Our analysis also allows us to provide a description of an infinite class of
new vacua along the Coulomb branch, where the RG flow alternates
energy ranges where the theory runs, and others where the theory is in
a superconformal phase. The borders between these subsequent regions
are described by  enhan\c{c}on-like rings and we naturally dub the
corresponding geometric structures enhan\c{c}on bearings.  We
provide the corresponding supergravity duals and show, both from the gauge  
theory and supergravity points of view, how such
vacua interpolate between the non-cascading and the cascading vacua.

The original supergravity solution of \cite{Bertolini:2000dk}, which
is the building block for all supergravity duals along
the Coulomb branch that we analyse, 
presents an unphysical repulsive region around the origin. Another 
distinctive property of $\calN=2$ models is the peculiar way in which
such a singularity is cured. Models with $\calN=2$ supersymmetry are not
confining, and the resolution of the IR singularity is associated to
the enhan\c{c}on mechanism \cite{Johnson:1999qt} which excises the
unphysical region giving back a singularity-free solution. The scale
at which the excision occurs depends on the dual gauge theory vacuum
one is studying \cite{Aharony:2000pp,Petrini:2001fk}, and therefore
the excised solutions will differ for different vacua. We work out
the enhan\c con mechanism for all gauge theory vacua mentioned above,
computing explicitly the warp factors of the excised solutions. It is
worth noticing that the way the enhan\c con mechanism works here is
qualitatively different from the original one discussed in
\cite{Johnson:1999qt}, since in the present case the enhan\c con shell
is not of real codimension one, {\it i.e.} it is not a domain wall: the modification
of the solution corresponds to an actual excision for the twisted
fields but not for the untwisted ones, most notably the metric and the RR 5-form field
strength. In turn, the corrected warp factor and 5-form depend on the
excised configuration of twisted fields and fractional branes dual to
the field theory vacuum under consideration. We find that around the 
origin the metric is free of singularities and the new solutions we 
find perfectly match, within the supergravity approximation, the 
dual gauge theory expectations.

The paper is organized as follows. In section \ref{sec:review cascading solution} we briefly 
recall the $\calN=2$ quiver gauge theory at the $A_1$ singularity, the
structure of its moduli space and that of the known supergravity duals,
both for the conformal and non-conformal models. In section 3 we
recall how the non-perturbative dynamics of the model can be studied through
Seiberg-Witten curves, and review the enhan\c{c}on mechanism. Section 4, which includes the main result 
of this work, is devoted to the analysis of 
the cascading vacua, while in section 5 we discuss the new class of 
vacua characterized by the presence of subsequent enhan\c con bearings. Finally, in section 6 
we work out the excision procedure and the corresponding warp factors for all the 
gauge theory vacua previously discussed. Conclusions, outlook and an 
appendix follow.


\section{D3 branes on the $\mathbb{C}^2/\mathbb{Z}_2$ orbifold and a cascading solution}
\label{sec:review cascading solution}

\begin{figure}[tn]
\centering
\hspace{1cm}
\includegraphics[width=.5\textwidth]{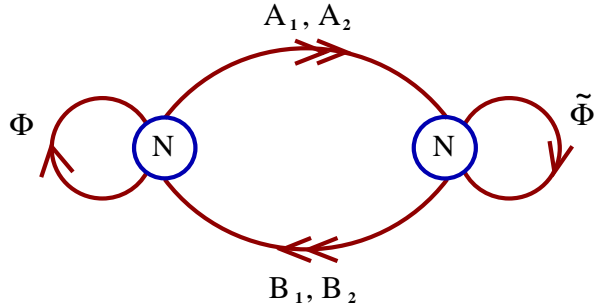}
\caption{\small Quiver diagram of the $U(N)_L \times U(N)_R~$ \Nugual{2} theory, in \Nugual{1} notation.
Nodes correspond to gauge factors, arrows connecting different nodes represent bifundamental chiral superfields
while arrows going from one node to itself represent adjoint chiral superfields.
\label{fig:N=2 matter content and quiver}}
\end{figure}

The low energy theory on $N$ D3 branes placed at the origin of the $\bbC \times \bbC^2/\bbZ_2$ orbifold is a
four-dimensional $U(N)\times U(N)$ \Nugual{2} gauge theory with two bifundamental hypermultiplets.
The field content is summarized in the quiver diagram of figure \ref{fig:N=2 matter content and quiver}.
The beta functions of both $SU(N)$ factors vanish, the diagonal $U(1)$ is decoupled, while the anti-diagonal $U(1)$
becomes free in the IR and gives rise to a global symmetry, the baryonic symmetry $U(1)_B$.

The classical moduli space agrees precisely with the possible configurations of regular and fractional D3 branes on
$\bbC\times \mathbb{C}^2/\mathbb{Z}_2$. In terms of \Nugual{1} superfields, the tree level superpotential
(dictated by \Nugual{2} supersymmetry) reads
\be
W = (B_1 \Phi A_1 - B_2 \Phi A_2) - (A_1 \tilde\Phi B_1 - A_2 \tilde\Phi B_2)~ ,
\ee
where contractions over gauge indices are implied.
The corresponding F-term equations are
\be
\Phi A_i - A_i \tilde\Phi = 0 ~~,~~ B_i \Phi - \tilde\Phi B_i = 0 ~~,~~ A_1B_1-A_2B_2 = B_1A_1-B_2A_2 =0 ~.
\ee
The holomorphic gauge invariant operators, which descend to local coordinates on the moduli space, are given by
traces of products of the operators $A_i B_j \equiv \varphi_{ij}$ and  $\Phi$ for the first gauge group, and
$B_i A_j \equiv \tilde\varphi_{ij}$ and  $\tilde\Phi$ for the second one.

The moduli space consists of several branches. First we have the so-called Higgs branches, where the hypermultiplets
obtain vacuum expectation values (VEV's). These VEV's result in the Higgsing of the quiver to a subgroup of the diagonal $U(N)$
gauge group, and the theory has an accidental \Nugual{4} supersymmetry in the IR. The Higgs branch has $(\bbC\times \bbC^2/\bbZ_2 )^N/S_N$
geometry, corresponding to the displacement of regular D3 branes in the full transverse space, up to permutations. Because of
\Nugual{2} supersymmetry, the K\"ahler metric on the Higgs branch is protected against
any quantum corrections. Next we have the Coulomb branch, on which the hypermultiplet VEV's vanish while the VEV's
for the two adjoint scalars can take arbitrary values: at a generic point on this branch, the surviving gauge group
is $U(1)^{2N}$. The Coulomb branch has the form $\bbC^N/S_N \times \bbC^N/S_N$, which corresponds to the displacement of
the two types of fractional D3 branes, each of them associated to one gauge factor, along the orbifold singularity
line. The quantum corrected metric on the Coulomb branch is exactly calculable thanks to Seiberg-Witten
theory \cite{Witten:1997sc}. Finally, there are mixed branches, where some hypermultiplet VEV's and some adjoint VEV's are turned on.

In the large $N$ and large 't Hooft coupling limit, the low energy superconformal $SU(N) \times SU(N)$ sector is better 
described by its type IIB supergravity dual \cite{Kachru:1998ys}. The  full Higgs branch is dual to a family of
supergravity solutions corresponding to D3 branes at arbitrary positions on the 6-dimensional transverse space,
\bea\label{metricAnstass}
ds^2 &= Z^{-1/2}\, \eta_{\mu\nu} dx^\mu dx^\nu + Z^{1/2}\, \delta_{nm} dx^m dx^n \\
g_s \,F_5 &= (1 + *) \, \dvol_{3,1} \wedge dZ^{-1} ~,
\eea
where $\mu,\nu = 0,\dots,3$, $m,n = 4,\dots,9$ and the orbifold identification  $\mathbf{x}=(x^m) \simeq ( \mathbf{\tilde x})
\equiv (x^{4,5}, -x^{6,7,8,9})$ is understood. $Z$ is a harmonic function of $\mathbf{x}$,
\be\label{warpHiggs}
Z = 4\pi g_s \alpha'^2 \sum_{j=1}^N \Big( \frac{1}{| \mathbf{x} -  \mathbf{x}_j|^4} + \frac{1}{| \mathbf{x} - { \mathbf{\tilde x}}_j|^4} \Big) ~.
\ee
The function contains the D3 branes and their images. Notice that
the total 5-form flux on $S^5/\bbZ_2$ at infinity is $N$.
The relation between the parameters  $x_j$ and the field theory moduli is $x_j = 2\pi\alpha' \phi_j$, where
$\phi_j$ is an eigenvalue of the VEV of some field. 
$\Phi$ and $\tilde\Phi$ are mapped to $x^4 + i x^5$, while $\varphi_{ij}$ are mapped to algebraic
coordinates $z_{ij}$ on $\bbC^2/\bbZ_2$,  such that $z_{12}z_{21} - z_{11}^2 = 0$ and $z_{22} = z_{11}$.
The supergravity axio-dilaton $\tau = C_0 + i\, e^{-\Phi}=C_0+\frac{i}{g_s} $ is constant,%
\footnote{We work in the string frame. Here $\Phi$ is the full dilaton, which is constant in all the solutions under
consideration, not to be confused with one of the adjoint chiral superfields. From now on we will rather use $g_s=e^\Phi$.}
as D3 branes do not couple to it. It is related to the field theory gauge couplings and theta angles by
\be
\tau = \tau_1 + \tau_2~~~ \text{where} ~~~ \tau_j = \frac{\theta_j}{2\pi} + \frac{4\pi i}{g_j^2}~,~~~~ j=1,2~.
\ee
In the following we will take $\tau=i/g_s$ unless otherwise stated.

As noticed in \cite{Aharony:2000pp}, for a generic point on the Higgs branch (and more generally on
any branch), the supergravity solution has large curvature. However, configurations where all the
branes are in big clumps have a good supergravity description, and configurations where only a small
number of branes are isolated are well described by probe branes in the background generated by the other branes.

The Coulomb branch of our \Nugual{2} quiver is described by fractional D3 branes along the orbifold
singularity. In this case supergravity solutions include a non-trivial profile for the twisted field fluxes. 
Indeed, fractional D3 branes source magnetically  the twisted scalar $c$ and by supersymmetry they also source
its NSNS partner, the twisted scalar $b$. This can be easily understood recalling \cite{Diaconescu:1997br}
that fractional D3 branes are D5 branes wrapped on the exceptional 2-cycle $\calC$ which lives at the orbifold
singularity. The twisted scalars are simply the reduction of the RR and NSNS 2-form potentials, $C_2$ and
$B_2$, on $\calC$. They can be organized in a complex field as
\be
\label{gamma def}
\gamma\equiv c+\tau b =c+ \frac{i}{g_s} b =
\frac{1}{4\pi^2\alpha'}\int_\calC \left( C_2 + \frac{i}{g_s} B_2 \right) ~,
\ee
while
\be
\label{G3 gamma}
G_3= F_3 + \frac{i}{g_s} H_3 = 4\pi^2\alpha'
\,d\gamma\wedge\omega_2
\ee
is the complexified 3-form field strength, where $\omega_2$ is a closed anti-selfdual $(1,1)$-form
with delta-function support at the orbifold plane, normalized as $\int_\calC \omega_2 = 1$.
Regular D3 branes do not couple to the twisted sector, hence the profile of $\gamma$ is affected solely by fractional
branes. The complex twisted scalar $\gamma$ is then subject to a two-dimensional Laplace equation in $\bbC$ with
sources at the positions of the fractional branes. Supersymmetric solutions \cite{Grana:2001xn} have primitive,
imaginary self-dual and $(2,1)$ $G_3$ flux, which implies that $\gamma=\gamma(z)$ is a meromorphic function of
$z = x^4 + i x^5$, such that $d\gamma(z)$ has simple poles at the locations of sources. For a bunch of $N$ fractional and $N$ anti-fractional%
\footnote{With some abuse of language, following \cite{Petrini:2001fk} we call `anti-fractional branes' D5
branes wrapped on $\calC$ with the opposite orientation, with some worldvolume flux through $\calC$ in order to 
preserve the same supercharges as the fractional branes.}
branes at positions $z_j$ and $\tilde z_j$, respectively, we have
\be
\gamma = \frac{i }{\pi}\, \Big[ \sum_{j=1}^N \log (z - z_j) - \sum_{j=1}^N \log(z - \tilde z_j) \Big] + \gamma^{(0)} ~.
\ee
Here $\gamma^{(0)}$ is an integration constant: its imaginary part sets the value of $b$ at large $|z|$ or in the theory
at the origin of the moduli space, while the real part does not really have a physical meaning in the dual theory because
of the presence of the axial anomaly, and we will set it to zero. The positions of the fractional branes $z_j$ and $\tilde z_j$
are classically identified with the eigenvalues $\Phi_j$, $\tilde\Phi_j$ of the field theory adjoint scalars. Corrections to
this identification arise at quantum level and will be discussed in the next section.

The holographic relations between the Yang-Mills couplings and theta angles and the supergravity fields are
\be \label{holographic relations}
\tau_1 + \tau_2 =\tau     \qquad\qquad\qquad     \tau_1 - \tau_2 = 2\gamma-\tau= 2 \Big[ c + \tau \Big( b - \frac{1}{2} \Big) \Big] ~,
\ee
but we will often set $\tau=i/g_s$.
In particular, when $b=0$ the imaginary part of $\tau_1$ vanishes and $g_1$ diverges, whereas for $b=1$ it is $g_2$
which diverges.%
\footnote{Actually $b \in [0,1]$ is the only range of validity of the formulas, because otherwise one would have negative square
couplings. As noticed in \cite{Klebanov:2000hb} and extensively discussed in \cite{Benini:2007gx, Argurio:2008mt},  when $b$ is
outside this range one has to perform a large gauge transformation to shift it to the interval where  (\ref{holographic relations})
can be applied.}
What we face in such cases is obviously a peculiar field theory, a SCFT with one divergent gauge coupling, in which instanton 
corrections dominate even in the large $N$ limit
\cite{Douglas:1995nw}, and about which not much is known. Although from the Seiberg-Witten curve analysis one does
not expect extra massless fields in general, the supergravity description is a very incomplete description for this
phase. When $c \in \bbZ$ as well, extra massless states do appear, and the theory enters a tensionless string
phase, as originally suggested in \cite{Witten:1995zh} from consistency of $T$-duality with type IIA string theory.

So far, we have only discussed the superconformal $SU(N) \times SU(N)$ theory,%
\footnote{From now on, we will often consciously forget the additional  $U(1)\times U(1)$ factor
which   decouples at low energies.}
 which has a well behaved UV limit and
whose stringy realization through AdS/CFT is unambiguous. However, what we are really interested in is the
non-conformal $SU(N+M) \times SU(N)$ gauge theory. This can be easily obtained through Higgsing from the superconformal
$SU(N+M) \times SU(N+M)$ theory, which can be engineered placing $N+M$ regular D3 branes at the origin of
the orbifold: taking $M$ VEV's of the second adjoint scalar to be at a scale $|z_0|/2\pi\alpha'$ produces an effective
$SU(N+M) \times SU(N) \times U(1)^M$ theory below $|z_0|$,%
\footnote{In the following, when speaking about scales we will often omit the $2 \pi \alpha'$
factor.} where the $U(1)$ factors are IR free and decouple. In the dual picture, this corresponds to placing $M$ anti-fractional
branes at, say, the roots of $\tilde z_j^M = -z_0^M$, while the other $N$ anti-fractional branes and $N+M$ fractional branes
sit classically at the origin. The twisted scalar in this configuration is then
\be\label{twisted scalar PRZ}
\gamma=\frac{i}{\pi}\, \log\frac{z^M}{z^M+z_0^M}+\gamma^{(0)}\;.
\ee
For the sake of simplicity, unless differently specified, in the following we will set the orbifold point value
$\gamma^{(0)}=\frac{i}{2g_s}$ \cite{Aspinwall:1996mn,Blum:1997fw}, so that in the UV $\tau_1=\tau_2=\frac{i}{2g_s}$.
In the large $M$ limit in which we work, \eqref{twisted scalar PRZ} can be traded for its limiting behavior
\be\label{twisted scalar PRZ large M}
\gamma=\begin{cases}
i\,\frac{M}{\pi}\, \log\frac{z}{z_0}+ \frac{i}{2g_s} \equiv i\,\frac{M}{\pi}\, \log\frac{z}{z_1} &\text{if} \quad |z|<|z_0|\\
\gamma^{(0)} &\text{if} \quad |z|>|z_0|
\end{cases}
\ee
where we set $z_1=e^{i\frac{\pi}{M}\gamma^{(0)}}\,z_0= e^{-\frac{\pi}{2 g_sM}}\,z_0$. 
Note that the twisted fluxes break the $U(1)$ isometry corresponding to rotation in the $z$-plane
to a discrete subgroup $\bbZ_{2M}$%
\footnote{Really, the cutoff fractional branes only preserve $\bbZ_M$, but 
this is irrelevant at large $M$ or not very close to those branes.}. This is dual to the breaking
of the $U(1)$ R-symmetry because of anomalies in the gauge theory \cite{Bertolini:2001qa}.

\medskip

The gauge invariant D3 brane charge (Maxwell charge) carried by the fluxes of the solution is proportional to the 5-form flux;
it is found by integrating the Bianchi identity in the absence of sources
$dF_5 = -H_3 \wedge F_3$ on the angular $S^5/\mathbb{Z}_2$ of radius $r$ and reads, for $r<\rho_0=|z_0|$,
\be
\label{5-form flux Matteo}
- \frac{1}{(4\pi^2\alpha')^2 } \int F_5 = N + \frac{g_s M^2}{\pi} \log \frac{r}{\rho_1} ~
\ee
with $\rho_1=|z_1|$.

We see from eqs.(\ref{twisted scalar PRZ large M})-(\ref{5-form flux
Matteo}) that, similarly to the Klebanov-Tseytlin (KT) solution
\cite{Klebanov:2000nc}, the solution
enjoys logarithmically varying B field and 5-form flux below the cutoff: this naturally suggests that the dual field theory
might enjoy a
cascading RG flow with subsequent infinite coupling transitions reducing the rank of the infinitely coupled non-abelian gauge
group by  $2M$ at scales $\rho_k= e^{-\frac{(2k-1)\pi}{2 g_sM}}\,\rho_0$, $k=1,\dots,l$, where $l\equiv [N/M]_-$%
\footnote{We denote by $[y]_\pm$ the ceiling and floor functions, namely the integers which better approximate $y$ from above
and below respectively.} \cite{Polchinski:2000mx}. This will be dealt with in section 4, where the $\calN=2$
cascading nature of the solution will be discussed in great
detail.

Before attacking this problem, though, we have to deal with another phenomenon, which always arises in supergravity solutions dual to
non-conformal supersymmetric gauge theories with eight supercharges. By analyzing the explicit form of the warp
factor, it was shown in \cite{Bertolini:2000dk} that the ten-dimensional metric obtained using \eqref{twisted scalar PRZ}, besides
the obvious singularity on the orbifold fixed plane, displays an
unphysical repulsive region near the origin, at a scale of order $e^{-\pi N/ g_sM^2} \rho_1$.%
\footnote{See appendix A of \cite{Billo:2001vg} for an analytic study of the warp
factor found in \cite{Bertolini:2000dk}.} One expects that, as suggested in \cite{Bertolini:2000dk}, an enhan\c 
con-like mechanism \cite{Johnson:1999qt} might be at work here, which excises the unphysical region rendering back a repulson-free solution.
We will show that this is indeed the case, discussing in the next section the specific way in which the enhan\c con mechanism
manifests in this context, and providing in section \ref{sec:excision} an excised and singularity-free solution.


\section{The enhan\c con and the Seiberg-Witten curve}
\label{sec:enhancon and SW}

The quantum corrections to the Coulomb branch constrain the (anti)fractional D3 brane positions, $z_j$ and $\tilde z_j$, 
in the gravity dual. The full
quantum corrected moduli space is exactly encoded in the full family of Seiberg-Witten (SW) curves
\cite{Seiberg:1994aj, Seiberg:1994rs}. The SW curves for the $\calN=2$ superconformal field
theory at hand were found in \cite{Witten:1997sc}. At the
classical level, the fractional brane positions $z_j$ and $\tilde z_j$
correspond to the eigenvalues of the VEV's of the adjoint scalars $\Phi$ and $\tilde \Phi$. In
the quantum theory this identification cannot survive because the VEV's
parametrize the moduli space and are unconstrained, whereas fractional brane positions are constrained.
That is, in the large $N$ limit one expects \cite{Aharony:2000pp} quantum
corrections and the consequent constraints on $z_j$ and $\tilde z_j$ to
be bound, because of supersymmetry, to a non-negative 5-form flux (that means non-negative enclosed D3-charge) for all
allowed configurations on the quantum moduli space, at least whenever the supergravity approximation is
valid.  This property is in fact at the core of the enhan\c{c}on mechanism.

Let us detail this point by first considering a simplified example. Consider
the theory discussed previously with $N=0$: this is an $SU(M) \times SU(M)$ superconformal
theory which can be engineered by $M$ regular D3 branes. Below the UV scale $|z|=|z_0|$, the theory
is effectively Higgsed to $SU(M)$ \Nugual{2} pure SYM
(plus IR free $U(1)$ factors).
The dual supergravity solution is the one in (\ref{twisted scalar PRZ large M})-(\ref{5-form flux Matteo})
with $N=0$, and it corresponds to the $M$ fractional branes classically at the origin.
The quantum moduli space can be studied with a good approximation by means of the SW curves for $SU(M)$ \cite{Klemm:1994qs,Argyres:1994xh}
\be \label{SW curve SYM}
y^2 = \prod_{a=1}^M (v-\phi_a)^2 + 4\Lambda^{2M} ~,
\ee
where $\Lambda$ is the strong coupling scale of \Nugual{2} $SU(M)$ SYM and $\phi_a$ are the eigenvalues 
of the adjoint scalar $\Phi$ parametrizing
a family of hyperelliptic curves in $\bbC^2=\{(v,y)\}$. The curves could also be written
in terms of gauge invariant symmetric polynomials. Classically ($\Lambda = 0$) the eigenvalues
$\phi_a$ coincide with the double branch points of (\ref{SW curve SYM}), and correspond to
the fractional brane positions on the $z$ plane in the gravity description. An elegant way
to see this is the following: type IIB string theory on the orbifold is T-dual to type IIA on
a circle (with coordinate $x^6$) with two parallel  NS5 branes along $x^0,\cdots, x^5$, separated
in the compact direction $x^6$ (see \cite{Giveon:1998sr} for a review). Fractional D3 branes are T-dual to D4 branes stretched
along $x^6$ between the two NS5's. The classical Coulomb branch is then given by all the possible
configurations of D4 branes on the plane $v=x^4+ix^5$.  The system can be further uplifted to
M-theory, where the NS5's and the D4's are just part of a single M5 brane. The M5 brane seen
as a Riemann surface is identified with the SW curve \cite{Witten:1997sc}. At the quantum level, the eigenvalues $\phi_a$ still
parametrize the whole moduli space (up to Weyl gauge identifications), but they no longer
correspond to double branch points nor fractional brane positions, strictly speaking. In
the perturbative regime of the theory, $|\phi_a| \gg |\Lambda|$, the branch points still
appear in pairs close to $\phi_a$: in the M-theory picture the D4 branes are inflated
into small tubes. As soon as the VEV's get into the non-perturbative region (at scales
comparable with $\Lambda$), the branch points get well separated and it does not make
much sense to talk about fractional brane positions anymore.

At the origin of the moduli space ($\Phi=0$), the hyperelliptic curve (\ref{SW curve SYM})
becomes $y^2 = v^{2M} + 4\Lambda^{2M}$, which has $2M$ separate branch points at
$v^{2M} = -4\Lambda^{2M}$. In the large $M$ limit, the branch points densely fill
a ring of radius $2^{1/M} |\Lambda|$. It is also possible to see that, adding a probe
fractional brane (in field theory terms, consider the $SU(M+1)$ theory with one additional
VEV $\phi$), in which case the SW curves are
\be
y^2 = v^{2M}(v-\phi)^2 + 4\Lambda^{2M+2} ~,
\ee
the probe can freely move in the semi-classical region outside the ring, but it cannot penetrate
it. For $|\phi| \gg |\Lambda|$, the two extra branch points are placed near $\phi$, with a
small separation of order $\Lambda(\Lambda/\phi)^M$, while the other $2M$ branch points are
still on the ring. As $|\phi|$ approaches $|\Lambda|$ and then goes to zero, the branch points
split and melt into the ring.

As anticipated, the dual string theory picture of this is the famous enhan\c con mechanism \cite{Johnson:1999qt}.
The tension of BPS fractional D3 branes is equal to their gauge invariant Maxwell D3-charge, which is  
$\gamma$
\be \label{tension fractional branes}
T_{n_f} = \frac{\mu_3}{g_s} \big| g_s \im \gamma+n_f\big| = \frac{\mu_3}{g_s} \big| b +n_f\big| ~,
\ee
where $n_f$ is the number of units of worldvolume flux on the exceptional 2-cycle $\calC$ (notice that neither $b$
or $n_f$ are gauge invariant, while their sum is).
This turns out to be proportional to the perturbative moduli space metric on the Coulomb branch of
the $SU(M)$ \Nugual{2} pure SYM theory%
\footnote{There is a matching with the perturbative result because in the large $M$ limit instanton
corrections are strongly suppressed, and abruptly show up at the scale $\Lambda$ \cite{Douglas:1995nw}.}
\cite{Bertolini:2000dk}. At the scale
$|\Lambda|= \rho_1$, $b$ vanishes and fractional D3 branes, which are wrapped D5 branes with no worldvolume flux, become
tensionless; below that scale they would be non-supersymmetric and they would feel a repulsive potential. Notice also
that the enclosed D3 brane charge would become negative for smaller scales, which could hardly be the case if fractional
D3 branes were at the origin. Moreover, a massive particle probe would experience an unphysical gravitational repulsion
close to the origin. The resolution of this puzzle is that fractional branes cannot be brought all at the same place, but
rather melt into a thin ring of radius $\rho_1$: the \emph{enhan\c con ring}. This changes the twisted fields distribution
in the geometry: inside the ring, $b=0$ (more generally it is integer), $c$ is constant, and there is no D3 brane charge.
The warp factor needs to be re-computed using the correct configuration of fractional branes and twisted field, and the
result is that the suspicious repulsive region disappears, as will be shown in section \ref{sec:excision}.

In some sense, the whole region defined by $b=0$ (more generally $b \in \bbZ$) behaves like a conductor:
D5 charges (recall that the D3 charge vanishes along with the tension inside the enhan\c con) are pushed to
the boundary and there is no field inside. We will call such a region the enhan\c con plasma. We
already noticed in section \ref{sec:review cascading solution} that the IR field theory dual to the
interior region is quite peculiar: it is a conformal $SU(N)\times SU(N)$ theory with one divergent gauge
coupling. However, in this particular case $N =0$ and the dynamics is trivial inside the enhan\c{c}on
plasma: $SU(M)$ is simply broken by instantons to $U(1)^{M-1}$.

As discussed in \cite{Petrini:2001fk}, exactly the same kind of behavior can be found in the most generic situation,
i.e. when $N \not = 0$ and the theory has product gauge group $SU(N+M) \times SU(N)$. Since the second gauge group
is not asymptotically free, one should embed the theory into the $SU(N+M) \times SU(N+M)$ conformal one, properly
Higgsed, as sketched at the end of Section \ref{sec:review cascading solution}. One can then exploit the power of
the Seiberg-Witten technology. In order to write down the SW curve, let us define the complex coordinate
\be
u= i\, \frac{x^6 + ix^{10}}{2\pi R_{10}} ~,
\ee 
which parametrizes the M-theory torus defined by the identifications $u \simeq u+1 \simeq u + \tau$. 
The complex structure $\tau$ is identified
with the type IIB axio-dilaton. Let us also define the parameter $q=e^{2\pi i \tau}$ and the coordinate
$t= e^{2\pi i u}$; note that $t\simeq qt$ on the torus.

For concreteness, let us stick again to the case of equal gauge couplings in the UV
CFT: $\tau_1 = \tau_2 = \tau/2$. In terms of the quasi-modular Jacobi $\theta$-functions 
\bea
&\qquad &
\theta_2(2u|2\tau) = \sum_{n=-\infty}^{\infty} q^{(n-\frac{1}{2})^2} t^{2n-1}\, \\
\theta_3(2u|2\tau) = \sum_{n=-\infty}^{\infty} q^{n^2} t^{2n}\, ,&\qquad &
\theta_4(2u|2\tau) = \sum_{n=-\infty}^{\infty} (-1)^n q^{n^2} t^{2n} ~,
\eea
the SW curve for the conformal theory can be written as \cite{Ennes:1999fb}
\be \label{SW curve f}
\frac{S(v) + R(v)}{S(v) - R(v)} = f(u|\tau), \quad \mathrm{with} \quad f(u|\tau)
\equiv \frac{\theta_3(u|\tau/2)}{\theta_4(u|\tau/2)} = \frac{\theta_3(2u|2\tau) +
\theta_2(2u|2\tau)}{\theta_3(2u|2\tau) - \theta_2(2u|2\tau)} ~,
\ee
or alternatively
\be \label{SW curve g}
\frac{R(v)}{S(v)} = g(u|\tau), \quad \mathrm{with} \quad g(u|\tau) \equiv \frac{f-1}{f+1} =
\frac{\theta_2(2u|2\tau)}{\theta_3(2u|2\tau)} ~.
\ee
Here $R(v) = \prod_{a=1}^{N+M} (v-\phi_a)$ and $S(v) = \prod_{a=1}^{N+M}(v-\tilde\phi_a)$ are degree $N+M$
polynomials whose zeros $\phi_a$ and $\tilde\phi_a$  are the eigenvalues for the adjoint scalars of the first
and second gauge group, respectively.

Following \cite{Petrini:2001fk}, let us choose a $\bbZ_M$-invariant configuration for the anti-fractional
branes Higgsing the CFT at large $|z|$ (i.e. large $|v|$ for the corresponding D4 branes), and consider the
origin of the moduli space of the low energy $SU(N+M) \times SU(N)$ theory,
\be\label{pt in ms cft 11}
R(v) = v^{N+M} \qquad\qquad\qquad S(v) = v^N(v^M - z_0^M) ~.
\ee
The $N$ common zeros of $R(v)$ and $S(v)$ factor out of the curve, without affecting the RG flow.
They correspond to $N$ D3 branes, whose moduli space is flat (apart from orbifold singularities when
several branes coincide) and not quantum corrected. We are then left to consider an $SU(M) \times SU(M)$
theory, spontaneously broken to $SU(M) \times U(1)^{M-1}$ at the scale $z_0$. Hence, if the IR dynamics
is not much affected by the UV Higgsing, as it is natural to expect, the low energy
physics should be similar to the enhan\c con mechanism previously discussed, but with $N$ leftover
regular D3 branes.

\begin{figure}[tn]
\centering
\includegraphics[width=.8\textwidth]{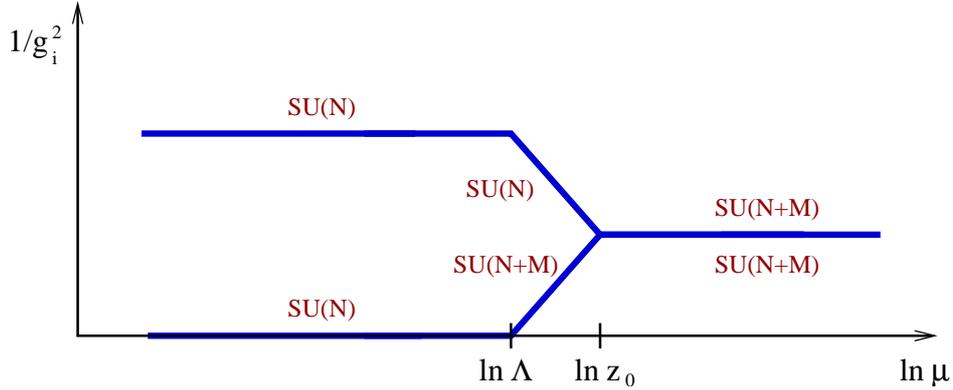}
\caption{\small RG flow of the theory at the enhan\c con vacuum (origin of the moduli space).
The low energy theory below $\Lambda$ is a peculiar one, with one formally diverging coupling. \label{fig:RG enhancon}}
\end{figure}

Let us give further evidence for the  above claim. As explained in \cite{Witten:1997sc}, we can extract the
running of the gauge coupling from the bending
of the two NS5 branes due to the unbalanced D4 branes tension. In the M-theory picture, the gauge couplings at a
scale $v$ can be extracted from the SW curve looking at the corresponding two values of $u$; we have that
\be
\Delta u =  \tau_1,\qquad \tau-\Delta u=  \tau_2 ~,
\ee
while the map between the type IIB twisted scalars ($c$, $b$) and the field theory couplings ($\tau_1$, $\tau_2$)
was given in (\ref{holographic relations}). In particular, the curve (\ref{SW curve f}) at the point
(\ref{pt in ms cft 11}) on the Coulomb branch reads
\be\label{SW curve for one enhancon}
 1 -2 \Big( \frac{v}{z_0} \Big)^M = f(u|\tau) ~.
\ee
One can check \cite{Petrini:2001fk} that in the UV regime $|v| > |z_0|$,
the theory is conformal with equal gauge couplings.
Comparing (\ref{SW curve for one enhancon}) with (\ref{SW curve SYM}), one can see%
\footnote{Notice that in the supergravity approximation, $g_s\rightarrow 0$ with $g_s N$ large, the parameter
$q=e^{2\pi i \tau}$ has exponentially small modulus $|q|= e^{-2\pi/g_s}$, allowing for a series
expansion of $f(t|q)$ in positive powers of $q$.}
that the dynamically generated scale is at $\Lambda= q^{\frac{1}{4M}}\,z_0$.
In the range $|\Lambda|<|v|<|z_0|$, the two gauge couplings are running with opposite $\beta$-functions
\be \label{beta functions}
\beta = \parfrac{}{\log{|v|}} \frac{8\pi^2}{g^2_{1,2}(|v|)} = \pm 2M ~.
\ee
For $|v|<|\Lambda|$ the gauge couplings are constant with $8\pi^2/g_{1,2}^2 = 0,\, 2\pi/g_s$ respectively. The
RG flow is sketched in figure \ref{fig:RG enhancon}. At the scale $\Lambda$, the gauge group is effectively
broken by instantons from $SU(N+M) \times SU(N) \times U(1)^M$ to  $SU(N) \times SU(N) \times U(1)^{2M}$,
the latter being conformal up to an IR free abelian sector.

Further information is gained from the computation of branch points of the SW curve, which correspond to double points of
the function $f(u|\tau)$: they are at $u_*=0,\, 1/2, \tau/2, (1+\tau)/2$ where $f(u_*|\tau)=
f_0,\, 1/f_0,\, -f_0,\, -1/f_0 $ respectively, and $f_0 = 1 + 4q^{1/4} + \calO(q^{1/2})$. The first set
is located at
\be
u = \frac{\tau}{2},\, \frac{\tau + 1}{2}: \qquad v \simeq v_h^{\pm} = z_0 e^{2\pi i h/M} \,
\Big[ 1 \pm \frac{2}{M} \Big( \frac{\Lambda}{z_0} \Big)^M \Big] \qquad h= 1,\dots, M ~.
\ee
These are almost double branch points, which correspond to the $M$ anti-fractional branes located near
$|z_0|$, corresponding to the VEV's of $\tilde\Phi$ we used to Higgs the conformal theory.
The second set is located at
\be
u=0,\, \frac{1}{2}: \qquad v \simeq v_k = 2^{1/M} e^{2\pi i k/2M} \Lambda \qquad k = 1, \dots , 2M ~.
\ee
These branch points correspond to $M$ fractional branes melted into an enhan\c con ring at scale $\Lambda$.

As in the pure SYM case, probe fractional branes can be studied on this background by means of the SW
curves for the $SU(N+M+1) \times SU(N+M+1)$ theory
\be 
\frac{R(v)}{S(v)} = \frac{v^M(v-\phi)}{(v^M- z_0^M)(v-\tilde \phi)} = g(u|\tau) ~,
\ee
where $\phi$ and $\tilde\phi$ parametrize the extra VEV for $\Phi$ and $\tilde\Phi$. The branch points corresponding 
to the eigenvalue $\phi$ (the fractional D3 probe)
can freely move outside the enhan\c con ring, but as they approach it and $\phi$ goes to  $0$, the two branch
points split and melt into the enhan\c con ring. The two branch points corresponding to the eigenvalue 
$\tilde\phi$ (the anti-fractional D3 probe) can instead
penetrate the enhan\c con ring; when this happens, they unchain two branch points from the ring which follow
them inside: an anti-fractional brane eats a melted fractional brane from the ring, forming a regular D3 brane
free to move everywhere.

From this analysis, one concludes that, no matter the value of $N$, the fluxes in
eqs. (\ref{twisted scalar PRZ large M}) and (\ref{5-form flux Matteo}) do
describe the physics of the $SU(N+M) \times SU(N)$ theory at the origin of its moduli space, provided
that they are excised at radius $\rho_1 \simeq |\Lambda|$ by an enhan\c con mechanism. The solution should
also be cut off at a radius $|z_0|$, or completed with $M$ anti-fractional branes, providing a conformal
$AdS_5$ UV completion. As already stressed, the warp factor needs to be recomputed in the presence of the correct
configuration of fractional branes and excised twisted fields. This will be done in section \ref{sec:excision}.

Notice, however, that the supergravity solution of eqs. (\ref{twisted scalar PRZ large M}) and (\ref{5-form flux Matteo})
does not seem to have any pathology below $\rho_1$, at least down to a scale of order
$e^{-\frac{\pi N}{g_sM^2}}\,\rho_1$, where the 5-form flux (\ref{5-form flux Matteo}) vanishes and the problematic
repulsive region starts. The question arises whether there is any field theory interpretation for such a solution,
suitably excised only at a radius 
\be
\rho_\mathrm{min}= \rho_{l+1}\equiv e^{-\frac{\pi l}{g_sM}}\,\rho_1\,\qquad \mathrm{with}\quad l \equiv [N/M]_- ,
\ee
 the smallest infinite coupling scale outside the region of negative D3 brane charge. As already
noticed, the  presence of a constant 3-form flux and the logarithmic running of the
 5-form flux strongly suggests a cascading behavior, as for the Klebanov-Tseytlin-Strassler $\calN=1 $
model \cite{Klebanov:2000hb, Klebanov:2000nc}, properly adapted to a $\calN=2$ setting. An interpretation of
the would-be \Nugual{2} RG flow that can be extracted from the supergravity solution in terms of some sort of Seiberg
duality cascade was in fact argued for in \cite{Polchinski:2000mx}, but the existence of an appropriate \Nugual{2}
duality had not been clarified, so far.
On the other hand, in \cite{Aharony:2000pp} the reduction of 5-form flux was interpreted as due
to a distribution of D3 branes and/or wrapped D5 branes. It was further
suggested that a suitable distribution of D3 branes only (Higgs branch)
could perhaps account for it. However, the latter proposal encounters
some problems in reproducing the running of gauge couplings and
decrease of nonabelian gauge group ranks that is suggested by the
supergravity solution.

Drawing on well established results about \Nugual{2} SQCD, we propose that there exist field theory
vacua, not at the origin of the Coulomb branch, which display a \emph{cascading} behavior. They are
dual to the solution in (\ref{twisted scalar PRZ large M}) and (\ref{5-form flux Matteo}), valid well below the
first infinite coupling radius $\rho_1$ down to some much lower scale, at most until the so-called true enhan\c con
scale $\Lambda_{min}=\rho_\mathrm{min}$, where the twisted fields are excised.
$\rho_\mathrm{min}$ is named the true enhan\c con radius since it is the scale at which the excision is performed.
All the higher infinite coupling scales, $\rho_j$ with $j=1,\dots,l$, will be called generalized enhan\c con radii  \cite{Aharony:2000pp}.

We provide a precise identification of these vacua in the next section. The excision of the twisted fields by means of the
enhan\c con mechanism and the disappearance of the naive singularity  will be discussed in section \ref{sec:excision}.
Depending on the field theory vacua one is studying, the
excision can take place at different scales, for instance at $\rho=\rho_1$, as in the vacua discussed
in this section, or at the bottom of the cascade, at the scale
$\rho=\rho_{\mathrm{min}}$, as for the cascading vacua to be discussed in
section 4.


\section{The cascading vacuum in field theory}
\label{sec: cascading vacuum field theory}

The perturbative RG flow of the $SU(N+M) \times SU(N)$ theory, given in (\ref{beta functions}), is such
that the largest group goes to strong coupling at a scale $\Lambda$.
The supergravity solution we are considering suggests that, in the dual vacuum, a mechanism effectively
reduces the gauge group to $SU(N-M) \times SU(N)$ below $\Lambda$, plus possible $U(1)$ factors.
This statement can be supported by a computation of Page charges in supergravity, in the gauge that gives
sensible field theory couplings (as extensively discussed in \cite{Benini:2007gx, Argurio:2008mt}). The
value of $b$, in the gauge in which $b \in [0,1]$, is found from (\ref{twisted scalar PRZ large M}) to be
\be
b = g_s \im \gamma = \frac{g_sM}{\pi} \log \frac{\rho}{\rho_1}  - \Big[ \frac{g_sM}{\pi} \log \frac{\rho}{\rho_1}  \Big]_- ~,
\ee
where  $\rho=|z|$. The D5 and D3 brane Page charges at radius $r$ are evaluated to be%
\footnote{The 3-cycle where the D5 charge integration is performed is the product of the exceptional 2-cycle $\calC$ and
an $S^1$ on the orbifold line. Since the intersection number is $(\calD, \calC)=-2$ where $\calD$ is the cone over the
3-cycle, the D5 charge is twice the number of wrapped D5 branes.}
\bea
Q_5^\mathrm{Page} &= - \frac{1}{4\pi^2\alpha' } \int F_3 = 2M \\
Q_3^\mathrm{Page} &= - \frac{1}{(4\pi^2\alpha')^2 } \int (F_5 + B_2 \wedge F_3) = N + M \Big[ \frac{g_sM}{\pi} \log  
\frac{r}{\rho_e} \Big]_- \;.
\eea
This shows that the non-abelian factors in the gauge group drop as $SU(N+M) \times SU(N) \to SU(N-M) \times SU(N)$ not only at
the first strong coupling scale $ \rho_1 = \Lambda_1 \equiv \Lambda$, but actually
at each generalized enhan\c{c}on, which occurs at a scale
\be
\label{scales}
\rho_k = \Lambda_k = e^{-\frac{\pi (k-1)}{g_sM}} \Lambda_1=  e^{-\frac{\pi (2k-1)}{2g_sM}} \rho_0 ~~k=1,\dots,l~~,
\ee
where recall that $l=[N/M]_-$ and we also set $N=lM+p$.
Finally, at $\Lambda_{l+1} \equiv \Lambda_{min}\equiv e^{-\frac{\pi l }{g_sM}} \Lambda_{1}$ there is a true enhan\c con ring
with $M$ tensionless fractional branes, and the non-abelian factors in the gauge group reduce according to
$SU(M+p)\times SU(p)\to SU(p)\times SU(p)$, with one infinite gauge coupling. Twisted fields have to be excised there so as
to avoid negative D3-charge in the interior region.

In passing let us stress, as in \cite{Aharony:2000pp}, that even though their dynamics takes place at arbitrarily low energies,
the possible additional $U(1)$ factors are described in the holographic setup by modes at a finite radius where the corresponding
fractional D3 branes lie.

In order to have an intuition on the strong coupling dynamics at hand, let us first focus on the first such generalized enhan\c{c}on,
which occurs at the scale $\Lambda_1=\Lambda$. This will clearly be  a prototype for any generalized enhan\c{c}ons. As already stressed,
at the scale $\Lambda$, the coupling of the largest gauge group
diverges (and instantonic  corrections dominate), while
the other gauge coupling reaches the value $g_\mathrm{min}^2 = 4\pi g_s$. As a first step toward the understanding of the precise
mechanism taking place, we can consider a corner of the parameter space of the gauge theory where $N g_\mathrm{min}^2 \to 0$. In
this limit, the gauge dynamics of the second factor decouples and it effectively becomes a global symmetry: the theory around
$\Lambda$ is simply $SU(N+M)$ SQCD with $2N$ flavors. Moreover, possible
VEV's for the smaller group adjoint scalar effectively behave as masses for the larger group hypermultiplets. In this case
we are out of the supergravity approximation but this analysis will give us some good insight. Hence, let us quickly
review some results about the moduli space of \Nugual{2} SQCD.

\subsection{One cascade step: \Nugual{2} SQCD}
\label{subsec: baryonic root SQCD}

The moduli space of $\calN=2$ SQCD \cite{Argyres:1996eh} with $N_c$ colors and $N_f$ flavors consists of a Coulomb branch and
of various Higgs branches. The Coulomb branch \cite{Hanany:1995na, Argyres:1996eh} is parametrized by the vacuum expectation
value of the adjoint scalar field $\Phi$ in the $\calN=2$ vector
multiplet,
\be
\Phi = \mathrm{Diag} (\phi_1, \dots , \phi_{N_c}) \qquad\qquad \sum_a \phi_a = 0 \;,
\ee
and is thus given by the $N_c -1$ dimensional complex space of $\phi_a$'s modulo permutations (Weyl gauge transformations).
The VEV's generically break the $SU(N_c)$ gauge group to its Cartan subgroup
$U(1)^{N_c-1}$. However, at special submanifolds where the Higgs
branches meet the Coulomb branch a non-abelian gauge symmetry
survives. Higgs branches can be divided into a baryonic branch and various non-baryonic branches (according to whether
baryonic operators acquire  VEV's or not); the corresponding intersections with the Coulomb branch were dubbed roots.%
\footnote{Issues related to the baryonic root of \Nugual{2} SQCD and the
mass deformation to \Nugual{1} were recently discussed in
\cite{Bolognesi:2008sw}.} Higgs branches are not quantum corrected, however their intersections among themselves and with the
Coulomb branch are modified at quantum level.

The SW curve describing the Coulomb branch for vanishing masses is \cite{Hanany:1995na,Argyres:1995wt}
\be
\label{SW curve SQCD}
y^2=\prod_{a=1}^{N_c} (x-\phi_a)^2 + 4\Lambda^{2N_c-N_f} x^{N_f} ~.
\ee
Nonbaryonic branches are labeled by an integer $1 \leq r\leq \min([N_f/2]_-,\, N_c - 2)$. The low energy effective theory at the roots are
the IR free or finite $SU(r)\times U(1)^{N_c-r}$ SQCD with $N_f$
hypermultiplets in the fundamental representation and charged under
one of the $U(1)$ factors. At special points along these submanifolds,
the SW curve shows that $N_c-r-1$ additional
massless singlet hypermultiplets arise, each one charged under one of
the remaining $U(1)$ factors. It is important that
there are $2N_c-N_f$ such vacua, related by the broken
$\bbZ_{2N_c-N_f}$ non-anomalous R-symmetry acting on the Coulomb branch.

The baryonic branch exists for $N_c \leq N_f$, and the baryonic root is a single point, invariant under the
$\bbZ_{2N_c-N_f}$ R-symmetry. Thus its coordinates on the Coulomb branch are%
\footnote{For $N_f > 3N_c/2$ there are other $\bbZ_{2N_c - N_f}$-invariant submanifolds. However the baryonic root is just one
point, and one can show that it in fact belongs to the submanifold (\ref{bar root}) \cite{Argyres:1996eh}.}
\be
\label{bar root}
\Phi_\mathrm{bb} = (\underbrace{0, \ldots ,0}_{N_f - N_c}, \phi\,\omega, \phi\,\omega^2, \ldots, \phi\,\omega^{2N_c - N_f} )~,
\ee
where $\omega = \exp\{2\pi i/(2N_c-N_f)\}$, for some value of $\phi$ (and $\phi=0$ classically). The gauge group is thus
broken to $SU(N_f - N_c) \times U(1)^{2N_c - N_f}$, which is IR free.%
\footnote{We assume $N_f<2N_c$ so that the microscopic theory is UV free. This bound is satisfied in the cascading quiver theory.}
 The requirement that a Higgs
branch originates from this root implies the presence of $2N_c - N_f$
massless hypermultiplets charged only under the $U(1)$ factors; this singles out a point in the submanifold
described by (\ref{bar root}). The result is $\phi = \Lambda$,
so that the SW curve takes the singular form
\be \label{SW curve bar root}
y^2=x^{2(N_f-N_c)} \big( x^{2N_c - N_f} +  \Lambda^{2N_c - N_f} \big)^2 ~.
\ee
The $x^{2(N_f - N_c)}$ factor
corresponds to an unbroken $SU(N_f - N_c)$ gauge group. The remaining $2(2N_c - N_f)$ branch points show up in coincident pairs,
located at  $x_k = \Lambda\, \omega^{k-\frac{1}{2}}$ with $k = 1,\dots,2N_c - N_f$,
corresponding to the $2N_c - N_f$ mutually local massless hypermultiplets.

The reason for this detour should be  clear by now: the non-perturbative dynamics at the baryonic root
preserves the same $\bbZ_{2N_c-N_f} = \bbZ_{2M}$ R-symmetry as the supergravity solution we are discussing,
and its low energy effective theory possesses an $SU(N_f - N_c)= SU(N-M)$ non-abelian gauge symmetry precisely matching the
numerology of the cascading interpretation. Hence, iterating the above procedure at the subsequent generalized
enhan\c{c}ons $\Lambda_k$ (where the higher rank gauge group  coupling diverges), it is
natural to propose the supergravity solution in (\ref{twisted scalar PRZ large M}) and (\ref{5-form flux Matteo}) (excised
only down at the true enhan\c{c}on $\rho_\mathrm{min}$) to be dual to a cascading $SU(N+M) \times SU(N)$ quiver gauge 
theory at subsequent baryonic roots of the strongly coupled gauge groups.%
\footnote{We should mention that a proposal for an \Nugual{2} cascade at the baryonic root has been alluded to
in \cite{Evslin:2004vs}, in the context of the M-theory realization of this elliptic
model.} In what follows, we will provide several checks for the validity of our proposal.

\subsection{The cascading vacuum in the quiver gauge theory}
\label{sec:SW curve cascading vacuum}

Let us now turn to the full quiver gauge theory $SU(N+M) \times
SU(N)$. The vacuum we propose as the dual of the full cascading
solution is a vacuum in which, at each step along the resulting
cascade, the largest of the two gauge groups goes to strong coupling
with a behavior analogous to the the baryonic root of SQCD. This
vacuum is invariant under the same non-anomalous $\bbZ_{2M}$ subgroup
of the R-symmetry as the supergravity solution we started
with. Moreover, not only has it the correct spontaneous symmetry
breaking pattern but also the correct RG flow, including the beta
functions and the separation of scales where the transitions occur, as
can be extracted from supergravity.

It is worth stressing that our vacuum does not sit exactly at the
baryonic roots, as there are no baryonic roots in the quiver theory
(see section \ref{sec:mass deformation} for an exception). However, it
does approximate them in the supergravity limit in which $q \to 0$,
which is the limit of interest to us.

\begin{figure}[tn]
\centering
\includegraphics[width=1.0\textwidth]{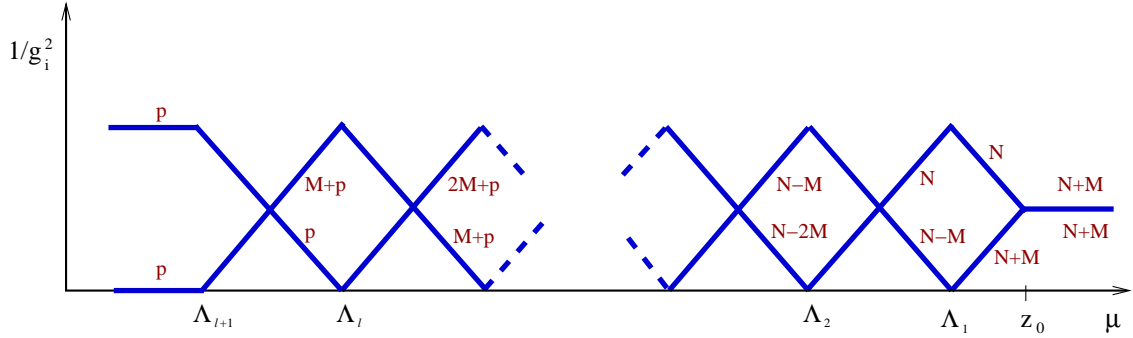}
\caption{\small RG flow of the theory at the cascading vacuum (taking $p=0$, for
definiteness). Here, as well as in figures \ref{fig:RG higgs} and \ref{fig:RG bearing}, 
the horizontal axis is logarithmic and we have omitted 
the $SU$ factors for the gauge groups, to avoid clutter. 
\label{fig: cascade}}
\end{figure}

Let us start for concreteness with an $SU((2K+1)M) \times SU((2K+1)M)$ conformal theory
in the UV and then break the gauge group to $SU((2K+1)M) \times
SU(2KM)$ by giving VEV's of order $z_0$ in a $\bbZ_{M}$-invariant
way to $M$ eigenvalues of the adjoint scalar $\tilde\Phi$. We choose
a vacuum in which, at each step of the RG flow, the most strongly
coupled group is at its baryonic root (in the $q \to 0$ limit). Let
us write the SW curve as $R(v)/S(v) = g(u|\tau)$ as in (\ref{SW curve
g}), where $u$ is the coordinate on a torus of complex structure
$\tau$. We choose the polynomials $R(v)$ and $S(v)$ of degree
$(2K+1)M$, as
\be
\label{cascading vacuum}
\begin{split}
R(v) &= v^M \prod_{j=0}^{K-1} (v^{2M} + q^{\frac{1}{2}+2j} \,
z_0^{2M}) \\ S(v) &= (v^M - z_0^M) \prod_{j=0}^{K-1} (v^{2M} +
q^{\frac{3}{2}+2j} \, z_0^{2M}) ~.
\end{split}
\ee
The polynomial $R(v)$ is related to the $SU((2K+1)M)$ group that
starts flowing toward strong coupling at the cutoff scale $z_0$,
whereas the polynomial $S(v)$ is related to the $SU((2K+1)M)$ group
which is spontaneously broken to $SU(2KM)$ there.%
\footnote{Very similarly, we can also describe a cascade with an
$SU(2KM) \times SU(2KM)$ UV completion: it amounts to putting the
cutoff and the vanishing eigenvalues in the same adjoint
field/polynomial in \eqref{cascading vacuum}, otherwise preserving the
structure of the polynomials.  Finally, the generalization to the
cascade with $N=lM+p$ can be achieved by multiplying $R$ and $S$ by
the same degree $p$ polynomial.} The eigenvalues of the two adjoint
scalar fields are put, in an alternating manner, at energies
corresponding to their subsequent strong coupling scales along the
cascade: in the limit in which the dynamics of the weakly coupled
group decouples at those scales, the vacua mimic the SQCD baryonic
root. In agreement with the cascading RG flow of the supergravity
solution, the hierarchy of  strong coupling scales is controlled by
$q=e^{2\pi i \tau}$. Because of the large $M$ limit, the running is led by the perturbative
beta functions except at the successive strong coupling scales,
where instantonic corrections sharply appear. This field theory
running can be explicitly checked either numerically using the exact
SW curve we wrote, or analytically by expanding the polynomials energy
range by energy range, in an effective field theory approach (see
Appendix \ref{check double points}). A plot of the resulting RG flow is shown in figure \ref{fig: cascade}.

We now move on to the study of the branch points of the curve. 
Recall that branch points are double solutions in $v$ at fixed $u$.
In the dual type IIA construction, a pair of coincident branch points at
$v$ corresponds to a D4 brane stretched between the two NS5's, while in 
type IIB it corresponds to a fractional brane  at position $z\simeq v$
on the orbifold singularity line. When the branch points are not in
pairs, the full M-theory description is needed, fractional branes are
no longer perturbative states in type IIB and their wavefunction is
spread over the whole $b \in \bbZ$ region \cite{Johnson:1999qt} (at
least in the large $M$ limit).

It turns out that the branch points for $u=0,\, 1/2$, up to
corrections of higher order in $q$, lie at
\be
\label{dblpoints 1}
v^M
\simeq \mp q^{n + 1/4}\, z_0^M \;, \quad n= 0, \dots ,K-1 \qquad\text{
and }\qquad v^M \simeq \mp 2 q^{K + 1/4}\,  z_0^M ~.
\ee
The former
class of points consists of $K$ sets of
$2M$ double points (which are double up to an accuracy discussed at the end 
of the next subsection), 
corresponding to the $K$ baryonic-root-like VEV's of
the first gauge group, whereas the latter are $2M$ well separated
branch points, corresponding to the true enhan\c con of the low energy
$SU(M)$ theory.  The branch points for $u = \tau/2,\, (\tau+1)/2$ lie
at
\be
\label{dblpoints 2} v^M = \mp q^{n + 3/4}\, z_0^M \;, n = 0,
\dots ,K-1 \qquad\text{ and }\qquad v^M = (1 \pm 2 q^{1/4})\, z_0^M ~.
\ee
The first class of points consists again of $K$ sets of $2M$ (almost) 
double points, corresponding to the $K$ baryonic-root-like VEV's of 
the second gauge group, while the second set of points are the almost paired 
branch points associated to semiclassical fractional branes at the cutoff 
scale $z_0$.

\subsection{The infinite cascade limit}
\label{sec:infinite cascade}

In this subsection we analyse the case of an infinite cascade, created
as the cutoff anti-fractional branes are sent to infinity. We are
interested in this limit for two main reasons: first of all, this
limit allows us to describe the field theory vacuum and the SW curve dual to
the infinite cascade solution of \cite{Bertolini:2000dk}, where there
are no cutoff anti-fractional branes; secondly, this infinite cascade
bears strong connections and similarities, that we will specify in the
following, with the Klebanov-Tseytlin-Strassler $\calN=1$ cascade
\cite{Klebanov:2000nc,Klebanov:2000hb}, which is necessarily unbounded
in the UV since fractional branes are stuck at an isolated conifold
singularity.

In order to properly define this limit, we should keep fixed the IR enhan\c con scale
$\Lambda_{min}$, as well as the generalised enhan\c con scales defined in~(\ref{scales}).
It is thus convenient to rewrite the two polynomials as
\be
\begin{split}\label{cascading vacuum 2}
R_K(v) &= v^M \prod_{j=1}^{K} (v^{2M}+q^{-2j}\Lambda_{min}^{2M}) \\
S_K(v) &= (v^M-q^{-\frac{1}{4}-K}\Lambda_{min}^M) \prod_{j=1}^{K}
(v^{2M}+q^{1-2j}\Lambda_{min}^{2M})~.
\end{split}
\ee
The limit of infinite cascade is formally $K\to \infty$.  Let us
define $x = (v/\Lambda_{min})^M$, obtaining the SW curve
\be
\label{SW_cascade}
T_K(x) \equiv \frac{R_K(v)}{S_K(v)} = \frac{x}{x
- q^{-1/4-K}}  \, \frac{\prod\limits_{j=1}^K (x^2 + q^{-2j})}{
\prod\limits_{j=1}^K (x^2 + q^{1-2j})} = g(u|\tau) ~.
\ee
Note that
\be
T_K(x) = \frac{x}{x - q^{-1/4-K}}  \, \frac{\prod\limits_{j=1}^K
(1 + q^{2j}x^2)}{ q^K\,\prod\limits_{j=1}^K (1 + q^{2j-1}x^2)}
\ee
converges pointwise as $K \to \infty$ for any fixed value of $x$ (possibly with poles) since
$|q|<1$, even though it does not converge uniformly.

We can then show that the approximate double points become exact at
any order in $q$ at large enough $|v|$ (i.e. the monopoles become
exactly massless in the upper reach of the cascade).  We will make use
of the following property of $g$ at its double points: $g \big( 0
\big| \tau \big) = -g \big( 1/2  \big| \tau \big) = 1/g \big( \tau/2
\big| \tau \big) = -1/g\big( (1+\tau)/2 \big| \tau \big)$. Moreover,
the value of the periodic function at these points is given by
\be
g_0(q)\equiv
g(0|\tau)=\frac{\theta_2(0|2\tau)}{\theta_3(0|2\tau)}=2q^\frac{1}{4}
\,\prod_{j=1}^\infty \frac{(1+q^{2j})^2}{(1+q^{2j-1})^2} ~.
\ee
Let us start with the branch points at $u=0,\, 1/2$ and $x =
-\epsilon\, q^{-n}$, where $n=1, \dots ,K$ and $\epsilon =\pm
1$. After some manipulations one gets
\be
T_K(-\epsilon\, q^{-n}) =
\frac{2\epsilon\, q^{1/4}}{(1+\epsilon\,  q^{1/4 + K -n})} \,
\frac{\prod_{j=1}^{\min(n-1,K-n)}
(1+q^{2j})^2}{\prod_{j=1}^{\min(n,K-n)} (1+q^{2j-1})^2} \,
\frac{\prod_{j=\min(n-1,K-n)+1}^{\max(n-1,K-n)}
(1+q^{2j})}{\prod_{j=\min(n,K-n)+1}^{\max(n,K-n)} (1+q^{2j-1})} ~.
\ee
Consequently, the equation $T_K(x)= \epsilon\, g_0(q)$ is solved
up to corrections $\calO(q^{2\min(n,K-n)+1})$,
$\calO(q^{2\min(n-1,K-n)+2})$ and $\calO(q^{1/4 + K - n})$. In
particular, in the case $K \geq 3n$ which is the lower part of the
cascade we get
\be
\frac{T_K(-\epsilon\, q^{-n})}{\epsilon\, g_0(q)} = 1 + \calO(q^{2n}) ~,
\ee
and the branch points we found are correct up
to $\calO(q^{2n})$. Similarly, for the branch points at $u=\tau/2,\,
(\tau+1)/2$ and $x = -\epsilon\, q^{-n+1/2}$ with $n=1, \dots,K$, we
get
\begin{multline}
T_K(-\epsilon\, q^{-n + 1/2}) = \frac{1}{2\epsilon\, q^{1/4} \, (1+s
q^{3/4 + K -n})} \times \\
\times \frac{\prod_{j=1}^{\min(n-1,K-n+1)}
(1+q^{2j-1})^2}{\prod_{j=1}^{\min(n-1,K-n)} (1+q^{2j})^2} \,
\frac{\prod_{j=\min(n-1,K-n+1)+1}^{\max(n-1,K-n+1)}
(1+q^{2j-1})}{\prod_{j=\min(n-1,K-n)+1}^{\max(n-1,K-n)} (1+q^{2j})} ~,
\end{multline}
and in particular, for $K \geq 3n$
\be
\frac{T_K(-\epsilon\,
q^{-n+1/2})}{\big( \epsilon\, g_0(q) \big)^{-1}} = 1 + \calO(q^{2n-1}) \;.
\ee
In order to show that these two sets of branch points are
double, we compute
\be
\frac{dT_K}{dx}(x) = T_K(x)\, \bigg\{
\frac{1}{x} + \sum_{j=1}^K \frac{2x}{x^2 + q^{-2j}} -
\frac{1}{x-q^{-1/4 -K}} - \sum_{j=1}^K \frac{2x}{x^2 + q^{1-2j}}
\bigg\}~.
\ee
One can show that $T_K'(-\epsilon\, q^{-n}) =
\calO(q^{n+1/2})$ and $T_K'(-\epsilon\, q^{-n+1/2}) = \calO(q^n)$ 
so that the points are double, up to sub-leading corrections (from numerical 
studies it seems that the corrections actually appear at some much 
higher order).

In a similar way, one shows that the non-double branch points at
$u=0,\, 1/2$ and $x = -2\epsilon$ (enhan\c con) are correct up to
$\calO(q)$, whereas the almost double ones at $u=\tau/2,\, (\tau+1)/2$
and $x =(1 + 2\epsilon\, q^{1/4}) q^{-1/4 - K}$ (cutoff) are correct
up to $\calO(q^{1/4})$.

Summarizing, our analysis shows that the SW curve \eqref{SW_cascade} for
the finite cascade has a well defined infinite cascade limit as we
send $K\to\infty$. We also evaluated to which degree the approximate
double points in the $q\to 0$ limit, appearing at all the strong
coupling scales except the smallest one, depart from being exactly
double; we find that in the infinite cascade limit the mass of the corresponding
monopoles goes to 0 for any value of $q$ as we consider higher and
higher scales up in the cascade, that is large $n$. Finally, only at
the bottom of the infinite cascade do we find equally separated double
points (in the $q\to 0$ limit), filling a true enhan\c con ring in the
large $M$ limit.

\subsection{Mass deformation}
\label{sec:mass deformation}

A not completely satisfactory feature of the cascading vacua we proposed
is that, although they preserve the $\bbZ_{2M}$ R-symmetry as the baryonic
root of SQCD, the extra light monopoles are strictly massless only in the
$q \to 0$ limit or for very large $n$. At finite $q$ and $n$, our vacua
are not really singled out as very special points in the moduli space.
Surely this is enough to our
purpose of finding the field theory vacua dual to the
supergravity solutions in (\ref{twisted scalar PRZ large M}) and
(\ref{5-form flux Matteo}). However, it will be useful to argue for
the existence of a cascading vacuum with exactly massless monopoles.

The task can be related to mass deformation of the \Nugual{2} theory
to \Nugual{1}, after the addition of a mass term for the adjoint
scalars
\be
W_\mathrm{mass} = \frac{m}{2} ( \Phi^2 - \tilde \Phi^2) ~.
\ee
In the case of \Nugual{2} SQCD, a mass deformation lifts the
moduli space and only the points on the Coulomb branch with $2N_c -
N_f$ extra massless monopoles survive, that is the baryonic root and
the $2N_f - N_c$ special points along the non-baryonic roots. The
reason is that in the dual M-theory picture a mass deformation
corresponds to a relative rotation of the two extended M5 branches
(NS5-branes in IIA), and this is possible only if the curve has genus
zero (because in the \Nugual{1} theory confinement breaks completely
the gauge group, and the genus of the M-theory/SW curve equals the
rank of the left over group). On the other hand, moduli space points
with massless monopoles are singular points where the genus of the
curve reduces, and a maximal number of them is needed to reach zero genus.

This suggests that a special point on the moduli space of the quiver
theory should be found after a mass deformation. There are two main
problems however. The first is that the cascading theory is obtained
from the conformal theory by spontaneous breaking at the cutoff $z_0$;
this is no longer a solution after mass deformation. A possible
solution is to consider an infinite cascade, as in the case of the
conifold theory.
From a more conservative point of view, one could consider an unstable
time-dependent field configuration with a finite cascade (with a large
number of steps) in which the VEV's for the spontaneous breaking are
very large but collapsing to zero. In this case the
dimensionless parameter controlling the time evolution of the field is
$\ddot\Phi / \Phi^3 = -(m/\Phi)^2$, which is in fact very small for
$\Phi \gg m$.  This mechanism would ``freeze'' the cutoff in this
limit. The other problem is that, unlike the SQCD case, after mass
deformation the far IR is $SU(M)$ \Nugual{1} pure SYM, whose $M$ vacua
break $\bbZ_{2M}$ to $\bbZ_2$.

These observations suggest that we should look for a genus zero SW
curve which breaks $\bbZ_{2M}$ to $\bbZ_{2}$, mimicking the curve for $SU(M)$, and
which describes an infinite cascade. Let us start from one of the $M$
genus zero curves of \Nugual{2} $SU(M)$ SYM: being of genus zero they
are parametrized by a complex coordinate $\lambda$, from which one
constructs two rational functions $v$ and $t$
\cite{Hori:1997ab,Witten:1997ep} 
\be v = \lambda +
\frac{\Lambda^2}{\lambda} \;,\qquad\qquad t=\lambda^M
\qquad\qquad\Rightarrow\qquad t^2 - P_M(v) t + \Lambda^{2M} = 0 ~, \ee
where $P_M(v)$ is a particular polynomial of degree $M$ in $v$. In the
following we will set $\Lambda=1$; then $P_M(v)$ is a Chebishev
polynomial \cite{Douglas:1995nw} 
\be P_M(v) = \Big[ \frac{v + \sqrt{v^2 - 4}}{2} \Big]^M +
\Big[ \frac{v - \sqrt{v^2 - 4}}{2} \Big]^M ~.  
\ee 
The genus zero
curve for the infinite cascade vacuum in the quiver theory is simply
obtained by wrapping the SYM curve on the torus,
\be\label{def Q infinity limit} 
Q =  \lim_{K \to \infty} Q_K
=\lim_{K \to\infty} \prod_{j=-K}^K F(q^j t, v) = 0 \qquad\qquad\text{with}\qquad
F(t,v) = t - P_M(v) + \frac{1}{t} ~,
\ee 
where $t = e^{2\pi i u}$. This definition  is mainly formal, as the infinite product above
does not converge. However its zero locus in $T^2\times \bbC$ (the
curve itself) is well defined, and it consists of the SYM curve
wrapped infinitely many times on the torus. It is clear that it has
genus zero (being non-compact, we mean that it is parametrized by
$\lambda$) and that it reproduces the correct IR behavior of $SU(M)$
SYM.

In order to make sense of it, and to check that it is the limit of
a sequence of SW curves for longer and longer cascades, with the correct
hierarchy of scales as expected from the RG flow at the baryonic
roots, we consider finite $K$ (eventually sent to $\infty$) and
rewrite the curve as
\be \label{tilde Q K}
\tilde Q_K = q^{K(K+1)}
f(q) \, Q_K = f(q) \Big(t - P + \frac{1}{t} \Big) \prod_{j=1}^K \Big(
1 - P t q^j + t^2 q^{2j} \Big) \Big( 1 - \frac{P}{t} q^j +
\frac{q^{2j}}{t^2} \Big) =0 ~,
\ee
where $f(q) = \prod_{j=1}^\infty (1
- q^{2j})(1 - q^{2j-1})^2$. The zero locus is the same as before, but
now the product converges as $K \to \infty$. Then, we define a sequence of
SW curves for $SU\big( (2K+1)M \big) \times SU \big( (2K+1)M \big)$
given by
\be \label{SWcurve LK}
\mathcal{Q}_K \equiv - \tilde R_K \,
\theta_3(2u|2\tau) + \tilde S_K \, \theta_2(2u|2\tau)=0 ~,
\ee
with the polynomials $\tilde R_K$ and $\tilde S_K$ chosen as
\bea \label{polynomials K}
\tilde R_K(v) &= P(v) \prod_{j=1}^K \big(
q^{2j}P(v)^2 + 1 - 2q^{2j} + q^{4j} \big) \\
\tilde S_K(v) &= q^{-1/4}
\big( 1-q^{K+1/4}P(v) \big) \prod_{j=1}^K \big( q^{2j-1}P(v)^2 + 1 -
2q^{2j-1} + q^{4j-2} \big) ~.
\eea
Using the identities
\bea
\theta_3
\big( 2u|2\tau \big) &= \prod_{j=1}^\infty (1-q^{2j}) \, \big( 1 + t^2
q^{2j-1} \big) \, \big( 1 +t^{-2} q^{2j-1} \big) \\ \theta_2 \big(
2u|2\tau \big) &= q^{1/4} (t + t^{-1}) \prod_{j=1}^\infty (1-q^{2j})
\, \big( 1 + t^2 q^{2j} \big) \, \big(1 + t^{-2} q^{2j} \big) ~,
\eea
one can explicitly verify that
\be
\tilde Q_K = \calQ_K \qquad\qquad \text{up to orders } \calO(q^{K+ 1/4}) ~.
\ee
Moreover, since the
polynomials $P_M(v)$ behave as $v^M$ for $v \gg 1$, one can check
that the hierarchy of scales of the cascading vacuum of subsection \ref{sec:infinite cascade} is
reproduced, up to IR corrections related to the different unbroken R-symmetries.

Let us comment on this result. Eq. (\ref{def Q infinity limit})-(\ref{tilde Q K}) defines a
genus zero curve with exactly double branch points for any value of $q$, which describes
a theory with
infinitely long cascade and exactly massless monopoles, dual to a specific type IIB supergravity solution
with no $AdS$ asymptotics. One could think
of realizing the theory by wrapping an M5 brane along the curve, and then
computing observables from it. However one could object that, unlike the \Nugual{1}
infinite KS cascade which makes sense as a field theory through holographic
renormalization \cite{Aharony:2005zr}, an infinite \Nugual{2} cascade probably
does not. The reason is that as we cascade down the IR-free $U(1)$ factors
accumulate, and an infinite cascade would require an infinite number of
photons at finite energies, which does not make much sense. Thus in (\ref{SWcurve LK})-(\ref{polynomials K})
we constructed a sequence of legitimate SW curves for any value of $K$, describing
larger and larger field theories with cascade which, although not having genus zero because of the UV cutoff, approximate
the genus zero curve (\ref{tilde Q K}) with arbitrary precision, for any value of $q$ and $M$. We could
compute observables in the sequence, getting in the limit the same answer as from (\ref{tilde Q K}). Therefore 
this procedure makes sense of the infinite cascade theory, in the sense that observables in finite sectors are 
insensible to the (possibly infinite number of) decoupled photons.

Eventually, notice that the sequence in (\ref{SWcurve LK})-(\ref{polynomials K}) contains
the finite $q$ corrections to the \Nugual{2} cascade that are required to have exactly massless monopoles 
and that were missing in (\ref{cascading vacuum 2}) because those were not visible in supergravity.

The mass deformation of this $\calN=2$ vacuum is particularly
interesting because it induces a flow from the cascading \Nugual{2}
theory to the \Nugual{1} Klebanov-Strassler (KS) cascade. This is
expected on the field theory side because the adjoint fields have to
be integrated out at the scale of the deformation mass parameter,
leaving the Klebanov-Strassler field theory at smaller energies.

This is clear also in M-theory. The genus zero SW curve we proposed is
the one of \Nugual{2} $SU(M)$ SYM, rewritten on the torus so as to create
an elliptic model. Similarly to the $M$ genus zero points on the Coulomb branch of
\Nugual{2} SYM which survive mass deformation and flow to the $M$ confining vacua
of \Nugual{1} SYM, the $M$ genus zero \Nugual{2} curves we proposed flow to the $M$
cascading vacua of the \Nugual{1} KS theory, whose IR is in fact \Nugual{1} SYM.

The rotated \Nugual{1} curve in the limit $m\to \infty$ is easily written. As before, we start rotating the SW
curve for $SU(M)$ SYM, exploiting the rational parametrization in terms of $\lambda$ \cite{Witten:1997ep}
\be
\left\{ \begin{aligned} v &= \lambda \\ t &= \lambda^M \\ w &= \zeta \lambda^{-1} \end{aligned} \right.
\qquad\qquad\Rightarrow\qquad \left\{ \begin{aligned} t &= v^M \\ vw &= \zeta \end{aligned} \right.
\ee
where the low energy strong coupling scale $\zeta=\Lambda_{\calN=1}^3=m \Lambda_{\calN=2}^2$ is kept fixed 
in the limit, and a suitable rescaling of variables is performed \cite{Hori:1997ab}.
The curve for the quiver theory is obtained by wrapping the curve on the M-theory torus: $0 = \prod_j (q^jt - v^M)$. 
After a rescaling to make the product converge, we get
\be
0 = (t - v^M) \prod_{j=1}^{K \to \infty} \Big( tv^M - q^j(t^2 + v^{2M}) + q^{2j} tv^M \Big)\, ,  \qquad\qquad vw = \zeta \;.
\ee
Note however that while in the \Nugual{2} case the M5 brane embedding can be interpreted as the exact SW curve for the 
field theory, which encodes the prepotential and the full dynamics, after breaking to \Nugual{1} this is no longer the 
case. The theory on the M5 brane reduces to the field theory of interest only when, for particular choices of the 
parameters, the unwanted modes are decoupled, and we refer to \cite{Witten:1997ep,deBoer:1997zy} for details.

It should be possible to reproduce this interpolating flow in
supergravity, so as to gain insight also on the K\"ahler data of these
$\calN=1$ vacua.  In particular, if the mass deformation is much
larger than the enhan\c con scale $\Lambda$, the solution should
interpolate to the Klebanov-Tseytlin (KT) solution (before chiral
symmetry breaking takes place in the IR). We leave the analysis of
such an interpolating solution, which should be performed along the lines
of \cite{Halmagyi:2004jy}, to the future.


\section{More supergravity duals: enhan\c con bearings}
\label{sec:bearing}

In this section we study other vacua of the  $SU(N+M)\times SU(N)$
theory, focusing on a class preserving the same $\bbZ_{2M}$ R-symmetry
as the supergravity solution of section 2. We will start from the non-cascading 
enhan\c con vacuum of section \ref{sec:enhancon and SW} and
gradually construct the cascading vacuum discussed previously by
pulling VEV's out of the  origin. In this process, we will observe new
nontrivial vacua, for which we will propose  novel type IIB dual
backgrounds.

Let us consider the following family of polynomials for the SW curves
of the $SU(N+M)\times SU(N+M)$ theory, parametrized by $\phi$ \be
\label{polynomials phi 1}
R(v)=v^{N-M}\,(v^{2M}-\phi^{2M}) \qquad\qquad S(v)=v^N \, (v^M -
z_0^M) ~.  \ee An overall $v^{N-M}$ factor (interpreted as $N-M$ D3
branes at the origin) decouples from the SW  curve (\ref{SW curve g}),
so that we will effectively reduce to the $SU(2M)\times SU(2M)$ case,
with \be
\label{polynomials phi 2}
R(v)=v^{2M} - \phi^{2M} \qquad\qquad S(v)=v^M \, (v^M - z_0^M) ~.  \ee
For $\phi=0$ we are at the origin of the moduli space of the $SU(2M)
\times SU(M)$ effective theory, where the enhan\c con mechanism takes
place.  We want to study the branch points of  the SW curve as we vary
$\phi$ continuously, in the supergravity approximation of small $q$,
so that $g_0(q)=2q^{1/4} + \calO(q^{5/4})$.  We will use the shorthand
notation  $\xi = v^M$ and define the enhan\c con scale $\Lambda =
2^{1/M} q^{1/4M} z_0$.

Let us first consider the branch points at $u=0,\, 1/2$, related to
the polynomial $R$. Depending on the value of $|\phi|$, we find:%
\footnote{We write the first corrections only when they are necessary
to split double branch points.}
\begin{itemize}
\item $|\phi^M|<|q^{1/4}z_0^M|$ \be \xi \,\simeq\, \pm \Lambda^M
\;,\qquad\qquad \xi \,\simeq\, \pm \Big( \frac{\phi^2}{\Lambda}
\Big)^M ~, \ee namely $2M$ equally separated branch points at the
enhan\c con ring and $2M$ equally spaced branch points at a ring  of
radius $|\phi^2/\Lambda|$;
\item $|\phi^M| > |q^{1/4}z_0^M|$ \be \xi \,\simeq\, \pm(1+\epsilon\,
q^{1/4})\, \phi^M\;, \qquad \epsilon=\pm1~, \ee namely $2M$ pairs of branch points on a circle of radius $|\phi|$.
\end{itemize}
The branch points at $u=\tau/2,\, (1+\tau)/2$ related to the
polynomial $S$, as long as $|\phi^M| < |q^{-1/4}z_0^M|$  which will
always be the case if $|\phi| < |z_0|$, are
\be
\xi \,\simeq\, (1 \pm
2q^{1/4})\, z_0^M\;, \qquad\qquad \xi \,\simeq\, \pm 4q^{1/2} \Big(
\frac{\phi^2}{\Lambda} \Big)^M \;,
\ee
namely $M$ pairs of branch points along a circle of radius $|z_0|$ and
$2M$ equally spaced branch points   on a ring of radius
$4^{1/M}q^{1/(2M)}|\phi^2/\Lambda|$.

In order to understand what the  supergravity solutions dual to these
vacua are, it will be useful to recall what are the BPS  fractional branes at
our disposal. They are obtained by wrapping D5 branes or anti-D5's
($\eta = \pm 1$ below, respectively)  on the exceptional 2-cycle with $n_f$
units of worldvolume flux. Their Wess-Zumino action reads
\be
 S_{WZ} = \eta\, \mu_3 \int_{M^{3,1}} \Big[ \tilde c_4 + (b+n_f) C_4 \Big] ~,
\ee 
where $\tilde c_4$ is a twisted potential dual to $c$. We will use
the notation D5$_{n_f}$ and $\overline{\text{D5}}_{n_f}$  for the
fractional branes with flux (recalling that $n_f$ is gauge dependent
while the D3-charge is gauge invariant).  The BPS objects are those
whose worldvolume flux ensures positive D3-charge $\eta(b+n_f)$, which
then equals the tension (\ref{tension fractional branes});  notice
that when the D3 charge exceeds one, we simply have a marginally
stable bound state of a fractional D3 brane with a number of regular D3
branes.

\begin{figure}[tn]
\begin{center}
\includegraphics[width=.75\textwidth]{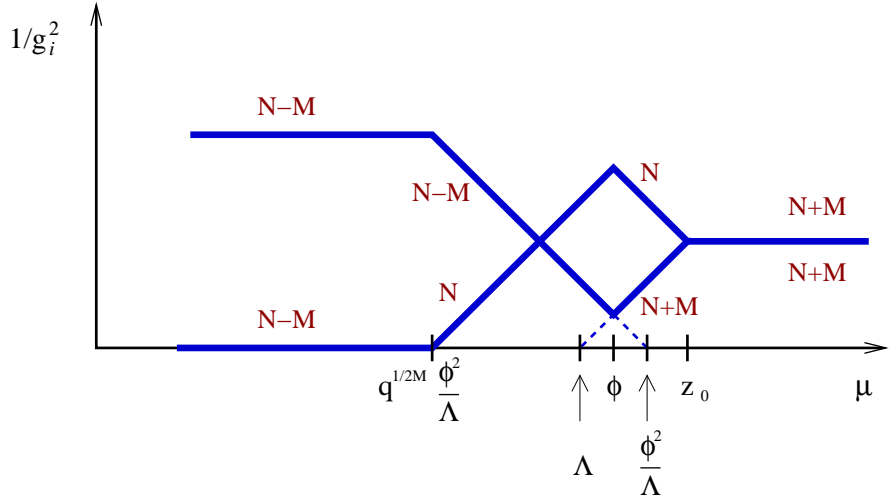}
\caption{\small RG flow of the theory at a vacuum with a perturbative
Higgsing at scale $\phi$. \label{fig:RG higgs}}
\end{center}
\end{figure}

The picture which stems from the branch points of the curve and from
the study of the RG flow is the following.

First, in the case  $|\Lambda| < |\phi| < |z_0|$, whose corresponding RG flow is depicted in figure \ref{fig:RG
higgs}, the theory is conformal in the UV, down to $z_0$ 
where $M$ eigenvalues of one adjoint scalar break the gauge group to
$SU(N+M)\times SU(N) \times U(1)^M$, triggering the RG flow. They
correspond to $M$ semiclassical $\overline{\text{D5}}_{-1}$'s in the
type IIB picture. At the scale $\phi$ there are $2M$ pairs of
branch points at the positions of the $2M$ VEV's of the other adjoint
scalar, which break further to $SU(N-M)\times SU(N) \times U(1)^{3M}$
and invert the RG flow. They correspond to $2M$ semiclassical D5's in
the geometry, which invert the twisted fluxes; in particular $b$
starts to grow as the radius decreases. At a lower energy scale
$q^{1/(2M)} \phi^2/\Lambda$ the $SU(N)$ coupling diverges, instantons
break the gauge group further to the conformal $SU(N-M) \times
SU(N-M)$ theory with one divergent coupling (times the $U(1)^{4M}$
factor), and we find $2M$  branch points equally spaced along a ring.
In type IIB, $b$ reaches the value 1 at the ring and there leaves $M$
tensionless $\overline{\text{D5}}_{-1}$'s smeared  over the enhan\c
con ring. It is possible to see by adding a
$\overline{\text{D5}}_{-1}$ probe that it cannot penetrate into the
interior, whereas a D5$_0$ can penetrate the enhan\c con ring,
unchaining a $\overline{\mathrm{D5}}_{-1}$ from it and making a D3 brane, which is free to
move inside.

\begin{figure}[tn]
\begin{center}
\includegraphics[width=5.5cm]{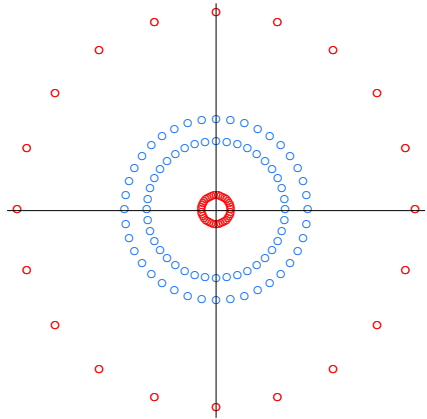}
\caption{\small Branch points of the $U(20) \times U(20)$ theory at a vacuum with one enhan\c con
bearing, a non-perturbative region  between two enhan\c con rings. Red (blue)
circles denote branch points related to the $S$ ($R$) polynomial.
\label{fig:bearing}}
\end{center}
\end{figure}

\begin{figure}[tn]
\begin{center}
\includegraphics[width=.85\textwidth]{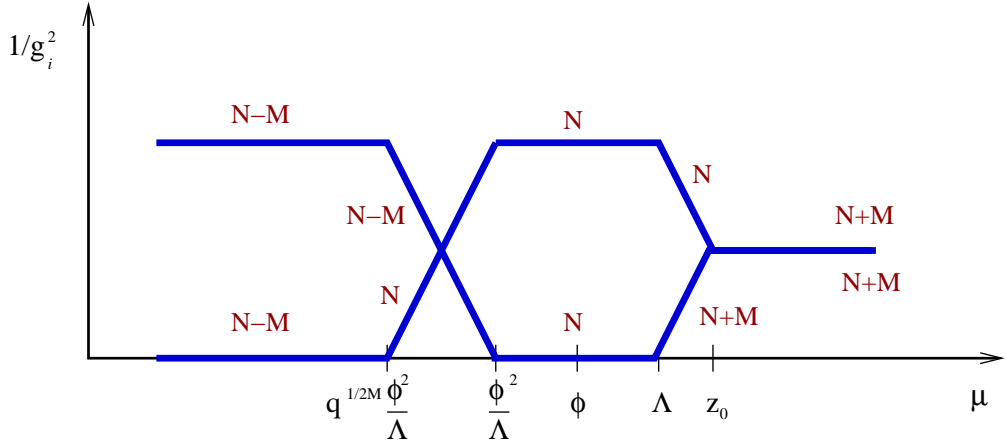}
\caption{\small RG flow of the theory at a vacuum with one enhan\c con
bearing. The theory is effectively the conformal $SU(N)\times SU(N)$ between the scales $\Lambda$ and $\phi^2/\Lambda$.
\label{fig:RG bearing}}
\end{center}
\end{figure}

There is a more interesting behavior in the case of $|\phi|<|\Lambda|$. If
$\phi=0$ we  are at the enhan\c con vacuum of Section
\ref{sec:enhancon and SW}. When $\phi$ does  not vanish, the branch
points follow the pattern of figure \ref{fig:bearing} whereas the RG
flow is the one depicted in figure \ref{fig:RG bearing}. As before, $M$ $\overline{\text{D5}}_{-1}$'s
are placed at the cutoff scale $z_0$.
From that scale downwards there is a flow with decreasing $b$ towards smaller radii, and an
enhan\c con ring with $2M$ equally spaced branch points at $\Lambda$,
where $b$ reaches 0 and $M$ tensionless D5$_0$'s are melted on the
ring. At lower energies the theory includes the conformal  $SU(N)
\times SU(N)$ factor with one divergent coupling: $b=0$ in the dual
supergravity solution, because of the $M$ fractional branes at the
enhan\c con ring. One could have expected that a new flow would start at a
scale $\phi$ because of the VEV's, but it does not: it actually starts
only at a lower scale $\phi^2/\Lambda$, where there are $2M$
additional equally spaced branch points; below this energy scale, the
gauge group with divergent coupling starts running towards weak
coupling again, whereas the other one runs towards strong coupling. We
enter a new perturbative regime, which ends with a final ring of
equally spaced branch points at scale $q^{1/(2M)}\, \phi^2/\Lambda$
where one gauge coupling diverges;  in the interior we find a new
conformal $SU(N-M) \times SU(N-M)$ sector, with one divergent
coupling, down to the IR.

We will call the ring at scale $\phi^2/\Lambda$ an anti-enhan\c
con. From the supergravity point of view it is indistinguishable from
a usual enhan\c con. However from the field theory point of view it is
quite peculiar: it represents instantonic effects that break the upper
conformal theory to a running one. These effects at the scale
$\phi^2/\Lambda$ are triggered by VEV's at the scale $\phi$: they take
some ``affine RG time'' to break the group; moreover this means that the effective
conformal theory must have some remnant of the scale
$\Lambda$. These issues deserve further investigations.

We dub the regions between enhan\c{c}on and anti-enhan\c{c}on rings, where 
$b \in \bbZ$ and the theory enjoys a superconformal phase, {\em enhan\c{c}on bearings}.

\medskip

It turns out that one can construct two different type IIB solutions
that describe this RG flow. The first one, say Higgsing-inspired (H),
by continuity with the case $|\phi|>|\Lambda|$ where a perturbative
Higgs mechanism takes place, interprets the ring of branch points at
$\phi^2/\Lambda$ as an anti-enhan\c con made of $M$ tensionless
D5$_0$'s (like the ones at $\Lambda$), which therefore force $b$ to
grow as the radius decreases, so that it remains bounded by 0 and
1. The innermost ring, placed where $b$ reaches 1,  is an enhan\c con
ring made of smeared tensionless $\overline{\text{D5}}_{-1}$.  In this
picture the D5$_0$'s ($\overline{\text{D5}}_{-1}$'s) are always
associated to the first (second) gauge group.

The second, say cascade-inspired (C), works by analogy with the
Klebanov-Tseytlin-Strassler \Nugual{1} cascade and interprets the ring
of branch points at $\phi^2/\Lambda$ as an anti-enhan\c con made of
$M$ tensionless $\overline{\text{D5}}_0$, and $b$ becomes negative at
smaller radii.  Then $b$ is monotonic, and the innermost ring at
$b=-1$ is interpreted as an enhan\c con ring made of $M$ tensionless
D5$_1$. This is the picture that matches with the solution in (\ref{twisted
scalar PRZ large M})-(\ref{5-form flux Matteo})  and which is usually
considered in the literature. The association between fractional branes and 
gauge groups is such that wrapped  (anti)D5 branes always correspond to the 
larger (smaller) gauge group.

Type IIB solutions like the two we are discussing here can be
explicitly constructed by excising and gluing twisted fields of the
solution in  (\ref{twisted scalar PRZ large M})-(\ref{5-form flux
Matteo}) (possibly generated by one  or the other kind of fractional
branes) and of a fluxless solution, with suitable sources
accounting  for the discontinuities at the glued surfaces, along the
lines of \cite{Johnson:2001wm}. As already stressed in
the case of the ordinary enhan\c con ring, this excision and gluing
procedure works for twisted fields, which are constrained to the
orbifold fixed plane. Instead, untwisted fields like the metric can
propagate also in the four dimensions of the orbifold, and must be
computed once the twisted fields and fractional brane configuration is
specified; this will be done in section \ref{sec:excision}.
It should be remarked that they turn out to be the same in
the two pictures.  One immediately realizes that all gauge invariant
quantities one could compute from the two solutions will give the same
answer, and in the field theory moduli space we have only one vacuum
to match with the two solutions. This suggests that an ambiguity must
be at work.

\begin{figure}[tn]
\centering
\includegraphics[width=.9\textwidth]{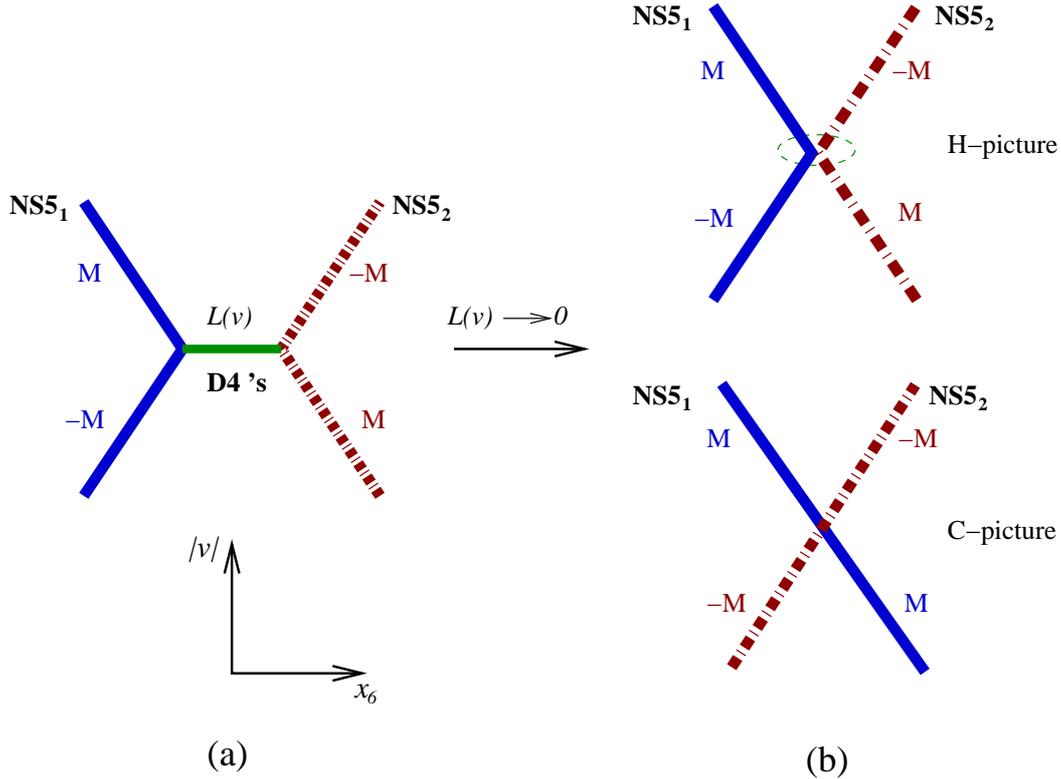}
\caption{\small IIA description and ambiguity. (a) a point of the moduli space where $2M$ D4 branes are 
stretched between two NS5 branes. (b) another point where the D4 branes have collapsed to zero length. 
In the H-picture we interpret the D4 branes as still present, providing bending tension and flux jump; 
in the C-picture, the D4 branes are simply not there. \label{fig:IIA picture}}
\end{figure}

The ambiguity is particularly apparent in the T-dual type IIA/M-theory
description. In type IIA, on each NS5-brane there is some worldvolume
$G_1 = dA_0$ flux. Space-time filling I3 brane intersections of
codimension two, where D4 branes end on an NS5 brane, are magnetic
sources for $A_0$; the flux $\oint G_1$ through any closed
path in the 2 dimensions of the NS5 worldvolume parametrized by $v$, in
which I3 branes are points, jumps by one unit whenever the path crosses
one of these points. In what follows we will consider circular paths
centered in the origin of the $v$ plane. One direction transverse to the NS5's, say $x^6$,
is compact of radius $R$ and the distance between the two NS5-branes
is $2\pi b\,R$. In figure \ref{fig:IIA picture}(a) we plotted the local
geometry around a ring where the perturbative Higgsing takes place as
in the RG flow of figure \ref{fig:RG higgs}: the NS5 on the left has
a flux $\oint G_1 = -M$ (in suitable units) below the stretched D4
branes, that jumps to $M$ above the D4's, while the opposite happens
to the NS5 on the right whose flux jumps from $M$ to $-M$. Along a
generalized enhan\c con ring $b$ is integer valued, so that the stretched D4's
are degenerate and the NS5's touch, as in figure \ref{fig:IIA
picture}(b). This interpretation leads to the H-picture in IIB: $b$ 
has a saw-shaped profile bounded by $[0,1]$ and there are $2M$
fractional branes of one kind in the enhan\c con bearing, $M$ on each
boundary. But the same IIA configuration can be equally well
interpreted as two NS5 branes that just cross, without any D4 branes
between them and without any jump in the flux. This leads to the
C-picture in IIB: $b$ is monotonic, and the bearing has fractional
branes on one side and anti-fractional on the other side, which cancel
their charge. In the type IIA picture there is clearly a single
configuration (dual to a single vacuum in field theory) which gives
rise to two pictures in IIB.

In type IIB, the ambiguity is related to S-duality: the duality
group $PSL(2,\bbZ)$ acts covariantly on the parameter space, whilst
the left over $\bbZ_2$ that acts as $(B_2, C_2) \to (-B_2, -C_2)$ 
and $(b,c) \to (-b, -c)$ on the twisted fields, is
gauged. The novel feature here is that the enhan\c con bearings are
domain walls on the $\bbC$ orbifold line, and the $\bbZ_2$ can act on
each domain separately.  At the same time, as already stressed, the
ambiguity does not affect the untwisted fields: $F_5$ and the warp factor
are the same in the two pictures, since they depend on the twisted
fields only quadratically in their field strengths; $B_2$ and $C_2$ are zero in the bulk.

\medskip

We can keep playing the same game of adding suitable VEV's, explained
so far in this section, to the newly found solutions, so as to
generate longer and longer RG flows with more and more transitions and
reductions of degrees of freedom. Of course the number of steps is at
most $[N/M]_-$. In this way we produce a class of vacua with a sort of
cascading behavior, with cascades of different lengths.

We conclude discussing the behavior of probes through the enhan\c con
bearing, as  extracted from the branch points of the SW curve with a
pair of VEV's added in  the perturbative regime outside the bearing,
and interpreting it in the C-picture (the other one is equivalent).
Consider first moving the VEV for the adjoint scalar of the gauge
group related to the branch points of the bearing, keeping the VEV for the
other adjoint fixed. As we decrease the VEV towards the outer enhan\c
con scale, the two branch points reach the ring and there split and
melt into it. Nothing happens until the VEV becomes smaller than the
scale of the inner anti-enhan\c con scale, when two branch points
escape from this ring, pair up and then continue their motion as
almost double branch points.  In the C-picture, this corresponds to a
D5$_0$ which melts at the outer enhan\c con, and later comes out of
the inner anti-enhan\c con as a $\overline{\text{D5}}_0$.  Similarly,
we can move the VEV for the adjoint scalar of the other gauge
group. The corresponding two branch points cross the outer enhan\c con
ring, unchaining two of its branch points. When they reach the inner
ring, they leave two branch points there and move on.  In
the C-picture, this corresponds to a $\overline{\text{D5}}_{-1}$ that
captures a D5$_0$  at the enhan\c con and becomes a D3-brane, free to
move inside the bearing; then it leaves a $\overline{\text{D5}}_0$ at
the anti-enhan\c con and becomes a D5$_1$ which is a minimal BPS
object in the region $b\in[-1,0]$ below the anti-enhan\c con ring.
This behavior of probes through the enhan\c con bearings in the case
of monotonic $b$ precisely  accounts for the non-trivial rearrangement
of minimal objects in BPS bound states claimed in \cite{Polchinski:2000mx}.

\subsection{Reconstructing the cascading vacuum at the baryonic roots}
\label{sec:reconstructing cascading vacuum}

We can now connect the enhan\c con bearing vacua discussed so far with
the cascading vacuum at the  baryonic roots of section \ref{sec: cascading vacuum field theory}.
Such a cascading vacuum has the  property
that all the complexified strong coupling scales along the cascade are
related by the  same hierarchy $q^{1/2M}$, which ensures that, at
least for $q \to 0$, the branch points pair up.

\begin{figure}
\begin{center}
\hskip -30pt
\includegraphics[width=\textwidth]{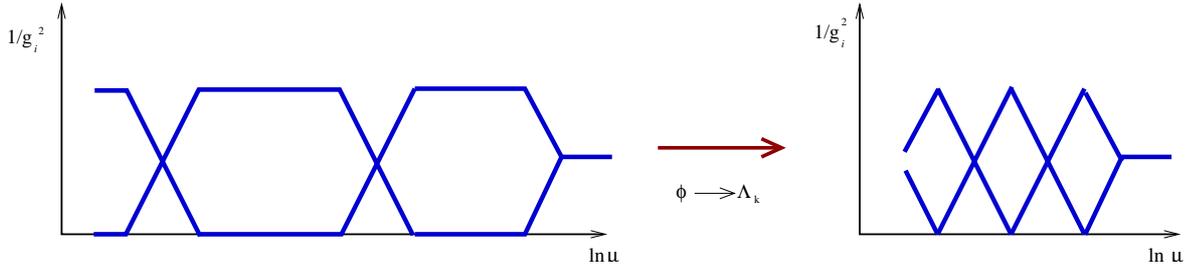}
\caption{\small In the limit where the thickness of the bearings is
sent to zero, one can reconstruct the cascading vacuum along
subsequent baryonic roots.
\label{fig:bear to zero}}
\end{center}
\end{figure}

We start from a vacuum with an enhan\c con bearing and send the
thickness of the bearing to zero sending $|\phi| \to |\Lambda|$
for the relevant strong coupling scale $\Lambda$. So doing, we end up
with a single circle at scale $\Lambda$ where $4M$ branch points lie,
$2M$ coming from inside and $2M$ coming from outside. For generic
phases of $\phi$, these branch points do not pair up, and on the type
IIB side we end up with a source term at the glued surface, accounting
for a discontinuity of $c$. If instead the phase of $\phi$ is suitably tuned,
branch points coming from the outer boundary and branch points coming
from the inner boundary of the bearing collide, hence forming double
branch points. Repeating the game with a vacuum with many enhan\c con
bearings, we can obtain the cascading vacuum along the baryonic roots sending the thickness of each bearing
to zero, see figure \ref{fig:bear to zero}.

In type IIB, as we reduce the bearing to zero thickness we make the
two smeared sources at the inner and outer boundaries of the bearing
coincide. Following \cite{Aharony:2000pp}, we call the resulting shell
a \emph{generalized enhan\c con ring}. In the H-picture, this is made
of $2M$ tensionless fractional branes, which account for the
$U(1)^{2M}$ factor left over by the gauge breaking. The presence of
the $2M$ massless hypermultiplets is more difficult to be claimed: one
could think of them as arising at the $2M$ points along the ring where
$\gamma \in \bbZ + \tau \bbZ$; however they should only be massless
for the correct tuning of the phase of $\phi$. Our belief is just that
the IIB supergravity description is incomplete at the enhan\c{c}on
bearings. On the contrary, in the M-theory description the mass of BPS
hypermultiplet states is given by the mass (proportional to the area) of M2 disks ending on the
M5 brane \cite{Witten:1997sc, Fayyazuddin:1997by, Henningson:1997hy, Mikhailov:1997jv} which is the same as the SW
curve; it is easy to see that the $2M$ double branch points
corresponds to massless hypermultiplets.

In the C-picture the generalized enhan\c con is made of $M$ fractional
and $M$ anti-fractional branes, both tensionless and
D3-chargeless. When the phase of $\phi$ is suitably chosen and the
inner and outer branch points coincide as we shrink the bearing, the
D5-charges locally cancel leaving the continuous supergravity solution
of Section \ref{sec:review cascading solution}; otherwise a source
remains accounting for the discontinuity of $c$, and one might think
of smeared  dipoles of fractional/anti-fractional branes. In this
picture the identification of the field theory modes is even subtler:
even when a perfect annihilation seems to occur, this cannot be the
case as the $U(1)^{2M}$ factor must still be there.

Let us conclude commenting on how the cascading vacuum at subsequent
baryonic roots naturally arises as the dual of the supergravity
solution of section \ref{sec:review cascading solution}.  Such
supergravity solution was constructed imposing rotational isometry on
the $\bbC$ orbifold line and without introducing any
source. Rotational isometry translates to $\bbZ_{2M}$ symmetry in
field theory, whilst absence of sources requires all the VEV's to be
at a strongly coupled scale. Among these vacua, only the cascading
vacuum in the C-picture avoid seeming discontinuities of $c$ (theta
angles) and $b$.

\subsection{More bearings: the enhan\c con plasma}

So far we have described a class of $\bbZ_{2M}$-symmetric solutions of
IIB supergravity, corresponding to vacua of  the dual field theory
with the same property, characterized by the presence of the enhan\c
con plasma in the shape  of fat rings (that we called enhan\c con
bearings). From a simple numerical inspection of the field theory
Coulomb  branch, one discovers that the enhan\c con plasma can take
quite different shapes (see for instance figure  \ref{fig: exotic
examples}). We give here a general characterization of such vacua, in
the large $N$ limit.

\begin{figure}[tn]
\centering \includegraphics[width=5.5cm]{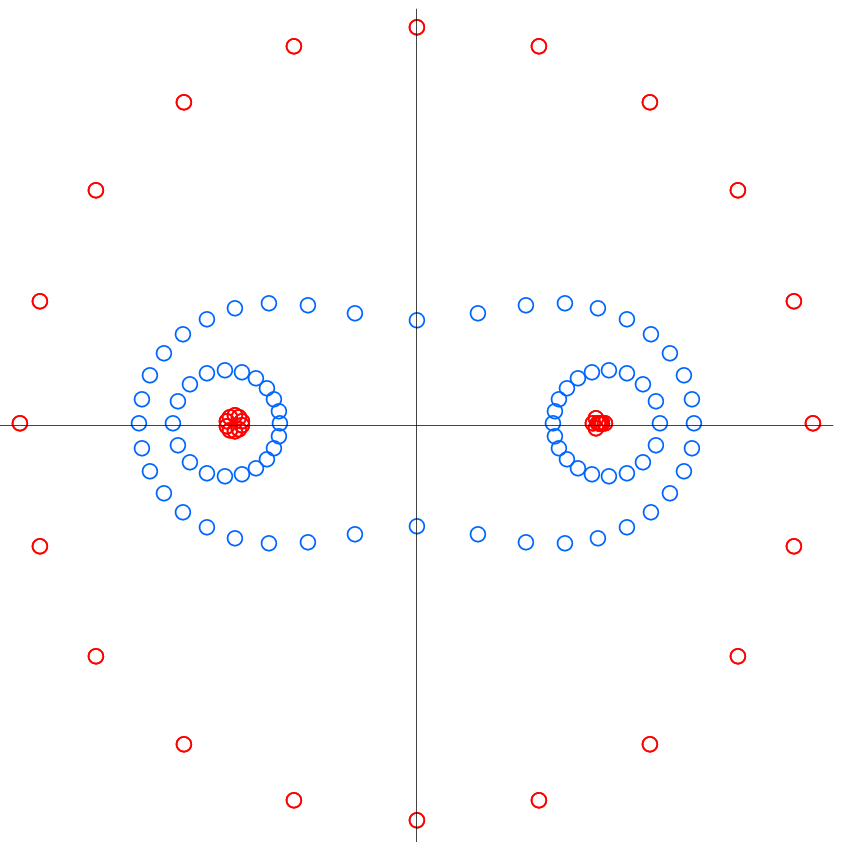}
\hspace{2cm} \includegraphics[width=5.5cm]{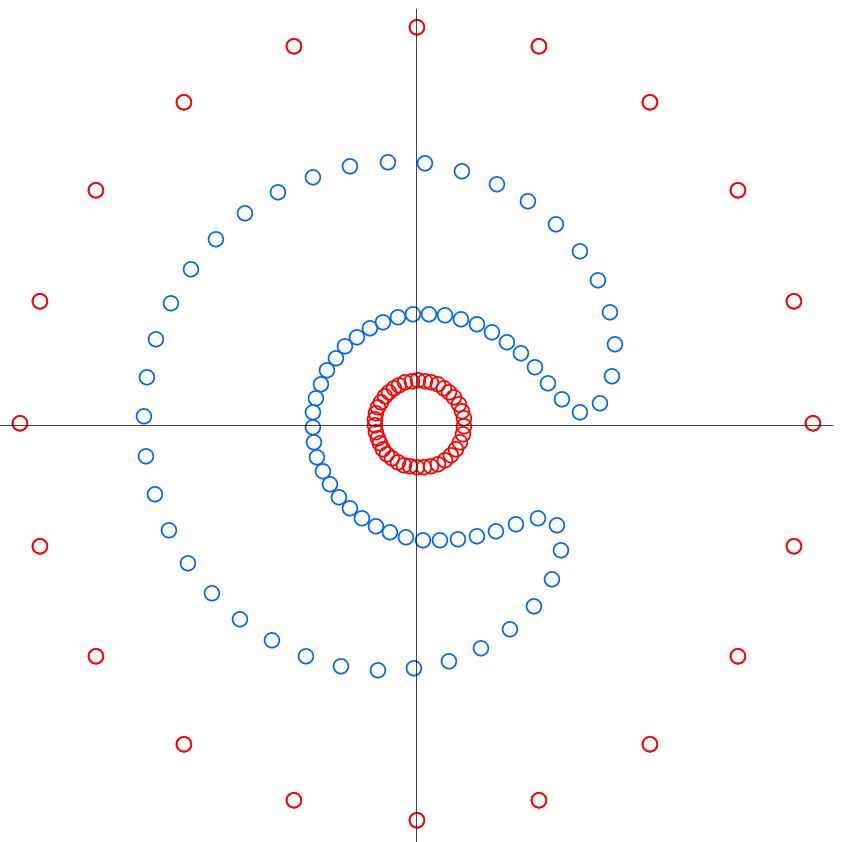}
\caption{\small Some vacua of the $U(40)\times U(40)$ in which the
enhan\c con plasma assumes exotic shapes. Left:  the plasma has two
holes. Right: the fat ring has broken into a horseshoe, disclosing the
inner region.
\label{fig: exotic examples}}
\end{figure}

We will show that from the point of view of IIB supergravity any
choice of the enhan\c con plasma domains, with the only  constraint of
charge quantization, leads to an actual solution and represents a field
theory vacuum.  For definiteness, we will study the $SU(N) \times
SU(N)$ conformal theory with $b = \frac{1}{2}$, spontaneously broken
to  non-conformal theories. Thus first of all we distribute some
number of anti-fractional branes in a circular ring of radius $\rho_0$
in the $\bbC$-plane orbifold singularity. Then we will arbitrarily
specify the enhan\c con plasma domains, without  any restriction on
the number of their holes and allowing nested domains.

The strategy to construct IIB supergravity solution is to solve for
the twisted potentials $b$ and $c$ first, and then for  $F_5$ and the
warp factor.

The enhan\c con plasma domains behave as conductors for the objects
carrying D5 charge, so that charges distribute themselves on the
boundaries and inside there are no fields: $b$ and $c$ are constant
with $b \in \bbZ$. Outside the plasma domains there are regions
$\calD_i$ where $b$ and $c$ are non-trivial. Consider one of these
regions, with its boundary given by a collection of curves
$\calC_{i,\alpha}$: there is one external curve $\calC_{i,E}$ while we
call the internal ones $\tilde \calC_{i,\alpha}$. The boundary
conditions in $\calD_i$ are that $b \in \bbZ$ on each curve
$\calC_{i,\alpha}$, and since we choose not to have generalized
enhan\c con rings nor tensionful fractional branes around (they both
can be obtained by sending to zero thickness an enhan\c con plasma
with fat ring or circular shape), up to gauge transformations and
picture ambiguity $b=1$ on $\calC_{i,E}$ and $b=0,1$ on $\tilde
\calC_{i,\alpha}$. The only exception is the outermost region
$\calD_E$ where $b=\frac{1}{2}$ on the external ring at $|z|=\rho_0$
and the branes are tensionful, while $b=0$ on
$\calC_{E,\alpha}$. Supersymmetry constrains $\gamma(z) = c +
\frac{i}{g_s} b$ to be a local meromorphic function, and after our
choice of boundary conditions actually holomorphic. To be precise,
$e^{-i\gamma}$ must be a holomorphic section of a $\bbC^*$
bundle. Rephrasing, we look for a harmonic real function $b$ with
fixed boundary conditions, and a local harmonic real function $c$
which satisfies the Cauchy-Riemann relations.

The problem of finding a harmonic function $b$ with prescribed values
on the boundaries $\calC_{i,\alpha}$  has one and only one
solution. It can be found by minimizing the functional 
\be
D[u]=\int_{\calD_i} |\partial u|^2 
\ee 
amongst all $u \in
C^{(1)}(\calD_i \setminus \bigcup_\alpha \calC_{i,\alpha}) \cap
C^{(0)}(\calD_i)$ with  $u\big|_{\calC_{i,\alpha}} =
b(\calC_{i,\alpha})$. A local harmonic function that satisfies the
Cauchy-Riemann  relation can be constructed as \be c(z) =
\frac{1}{g_s} \int_{p_0}^z \big( \partial_y b\, dx - \partial_x b\, dy
\big) \;, \qquad\qquad g_s\, dc = \partial_y b\, dx -  \partial_x b\,
dy = -*db ~, 
\ee 
where $z = x + i y$, $p_0$ is an arbitrary reference
point and $*$ is constructed with the flat metric on $\bbC$. Notice
that  $*d\gamma = -i\, d\gamma$. $c(z)$ turns out to be a multi-valued
function with monodromies which give a representation  of the homology
group of $\calD_i$. However, as long as the twisted field-strength
$dc$ is concerned, this is globally defined.

As we will explain in section \ref{sec:excision}, the warp factor is obtained 
by solving a Poisson equation (\ref{EqWarpFactor}) on the orbifold $\bbC^2/\bbZ_2 \times \bbC$, 
with two kinds of D3-charge source terms, both localized along the orbifold line. 
One is proportional to $|\partial_z \gamma|^2$  and comes from
the twisted fluxes. The other one is localized on the tensionful
fractional branes in the external ring and  represents their D3-charge. In general 
the brane density per unit length $\omega$ along
the ring is not constant but rather given by 
\be \label{density unit
length} \omega = - \frac{1}{2} \re\, \partial_t \gamma ~, 
\ee 
where the
derivative is taken tangent to the boundary. This comes from the
Bianchi identity $dF_3 \sim \delta^{(4)}_{D5}$.  On the boundaries of the
enhan\c con plasma domains there are fractional branes too with
density (\ref{density unit length}),  but they are tensionless as $b
\in \bbZ$ inside. Thus the only contribution of the latter kind comes
from the circular cutoff  ring at $|z|=\rho_0$. We do not go into
further details here, as the computation of the warp factor is fully
explained in section  \ref{sec:excision}. What matters is that there
is always one and only one solution normalizable at infinity. The
5-form flux is  then given by: $g_s F_5 = (1+*) \dvol_{3,1} \wedge
dZ^{-1}$.

So far we showed that for any choice of the enhan\c con plasma
domains, we can in principle solve the supergravity equations.
The last constraint is the D5-charge quantization, which amounts to
the monodromy of $c(z)$ being quantized 
\be \label{dc quantization}
\oint dc \in 2 \bbZ ~, 
\ee 
or in other terms $e^{-i\pi \gamma}$ being
a single-valued function. A basis of monodromies is given by $\Delta
c(\tilde \calC_{i,\alpha})$  on the internal boundaries $\tilde
\calC_{i,\alpha}$, and the integral is the total number of fractional
branes on them. As the  solution only depends on the choice of the
enhan\c con plasma boundaries (and the value of $b$ on them), (\ref{dc
quantization})  descends to a constraint (in fact the only one) for
them.

The total D3-charge of the system is then easily determined. The
contribution from the fluxes in all the regions $\calD_i$ is 
\be
\label{D3 charge fluxes} Q_3^{flux} = \frac{1}{2}\sum_i \int_{\calD_i} dc\wedge
db = -\frac{1}{2} \sum_i \sum_\alpha \int_{\calC_{i,\alpha}} b \, dc = -\frac{1}{2}
\sum_{i,\alpha}  b(\calC_{i,\alpha}) \, \Delta c(\calC_{i,\alpha}) ~.
\ee 
The contribution from the anti-fractional branes on the external
cutoff ring can be read from (\ref{density unit length})  to be:
$Q_3^\mathrm{cutoff} = -\frac{1}{4} \oint_{\rho_0} dc$, because
$b=\frac{1}{2}$ there. Since the external ring is  the external
boundary $\calC_{E,E}$ of the outermost region $\calD_E$, this
contribution can be added to (\ref{D3 charge fluxes})  by formally
considering $b(\calC_{E,E}) = 1$ instead of $1/2$. Notice that
(\ref{D3 charge fluxes}) is gauge and picture  invariant. However, for
our choice of gauge and picture the total charge is 
\be 
Q_3^{total} =
\sum_{i,\alpha} \big( 1 - b(\tilde \calC_{i,\alpha}) \big) \, \Delta
c( \tilde\calC_{i,\alpha}) \equiv N ~, 
\ee 
where we used that
$\calC_{i,E} = -\sum_\alpha \tilde \calC_{i,\alpha}$ in homology, and
$b(\calC_{i,E})=1$.  This expression counts the number of fractional
(as opposed to anti-fractional) branes. And in fact the solution we
constructed  is dual to a vacuum of the $SU(N) \times SU(N)$
theory. It is clear that if we want to embed this vacuum in a larger
theory, we  can simply add regular D3 branes.

Summarizing, we have shown that any choice of enhan\c con plasma
domains, up to the charge quantization constraint, gives rise  to a
solution of IIB supergravity with sources. Taking the limit of zero
thickness, we can also include generalized enhan\c cons  and isolated
bunches of fractional branes; bunches of regular branes are easily
included as well. Each of these solutions is dual to a vacuum on the Coulomb branch
of the $SU(N) \times SU(N)$ SCFT. Even though we cannot
be more specific about the exact map (it should be worked out
by computing operator VEV's holographically), 
this huge class of solutions helps in covering the
moduli space of the dual field theory.

\section{Excisions, warp factors and the cure of repulson singularities}
\label{sec:excision}

In this section we take into account the excision of twisted fields
inside the enhan\c con ring and bearings and work out the correct warp
factor for a quite general rotationally symmetric configuration of
fractional branes, which will be useful to describe the enhan\c con
vacuum of section \ref{sec:enhancon and SW}, the cut off cascading
vacuum of section \ref{sec:review cascading solution}, the infinite
cascade vacuum of section \ref{sec:infinite cascade} and the vacua
with rotationally symmetric bearings of section \ref{sec:bearing}.

We stress once again that consistency of the configuration of
fractional branes, in agreement with the dual field theory picture
encoded in the SW curve, implies an excision of the naive twisted
field solution at enhan\c con rings. Unlike the situation of
\cite{Johnson:1999qt}, where there is an enhan\c con shell of
codimension 1 in the non-compact part of the internal geometry, here we
face enhan\c con rings having codimension 1 only for the twisted
fields which are constrained to live on the orbifold plane, but not
for the bulk fields which propagate also in the four additional
dimensions of the orbifold. Consequently, the usual excision of
\cite{Johnson:1999qt,Johnson:2001wm} works for twisted fields but not
for untwisted fields; in particular, the warp factor has to be
computed once and for all, once the correct configuration of
fractional branes and twisted fields describing some gauge theory
vacuum is specified.

The equation which determines the warp factor $Z$ follows from the
modified Bianchi identity for $F_5$ in the presence of sources at the
locations of tensionful (anti-)fractional branes; it is a Poisson's
equation which reads \cite{Bertolini:2000dk}
\begin{equation} \label{EqWarpFactor}
\Delta_6 Z + (4\pi^2\alpha')^2 g_s^2\,|\partial_z \gamma|^2
\,\delta^{(4)}(\vec{x})+ 2(4\pi^2\alpha')^2 g_s\,\sum_i
Q(\mathbf{x}_i) \delta^{(6)}(\mathbf{x}-\mathbf{x}_i)=0 \;,
\end{equation}
where $\Delta_6$ is the 6-dimensional Laplacian and
$\mathbf{x}=(\vec{y},\vec{x})$ a 6-dimensional vector,
$\vec{y}\equiv(\re z,\im z)=(x^4,x^5)$ being a vector on the orbifold
fixed plane $\mathbb{R}^2$ and $\vec{x}=(x^6,\dots,x^9)$ being a
vector in the covering space $\mathbb{R}^4$ of the orbifold.  In the
previous formula, $Q(\mathbf{x}_i)$ is the gauge invariant D3 brane
charge of a regular or (anti-)fractional D3 brane placed at
$\mathbf{x}_i$, which depends on the object and on the value of fields
at its position (in the case of fractional branes). The sum runs over
all tensionful fractional D3 branes as well as regular D3 branes along
with their images.

We will first consider $M$ tensionless fractional branes melted in an
enhan\c con ring of radius $\rho_e$ in the fixed plane parametrized by
$z$, together with $M$ `cutoff' anti-fractional branes at the $M$
roots of $z^M=-z_0^M$, which are used to Higgs the  conformal UV
theory at the scale $\rho_0=|z_0|$. Here and in the following,
$\rho_e$ is the scale at which the excision should be performed and
its actual value depends, case by case, on the vacuum one is actually
considering. We will also impose that the total gauge invariant D3
brane charge of the configuration be $N+M$, adding regular D3 branes
at the origin when needed, so that the dual gauge theory is
$SU(N+M)\times SU(N+M)$ in the UV.  Using the freedom of shifting the
axion $b$ by an integer via a large gauge transformation, we will also
set $b(\rho)=0$ for $\rho <\rho_e$. Finally, we will be general and
place the cutoff anti-fractional branes at a scale $\rho_0$ such that
$b(\rho_0)$ can acquire any positive value; 
the gauge invariant D3 brane charge supported by each of the
anti-fractional branes is therefore
\begin{equation}
\label{charge_cut-off_branes}
- n_f - b(\rho_0)=[b(\rho_0)]_+ - b(\rho_0) ~.
\end{equation}
In other words,  these anti-fractional branes are D5 branes wrapped on
$-\calC$, with $-[b(\rho_0)]_+$ units of worldvolume flux on it.
Being in the large $M$ limit, we can safely approximate the cutoff
anti-fractional branes with a ring.

The warp factor gets different contributions. First of all, if there
are some regular D3 branes at the origin, they source the usual term
according to \eqref{warpHiggs}. Secondly, the $M$ cutoff
anti-fractional branes, because of their tension, contribute the
following term in the ring approximation
\begin{equation}\label{warp_ring}
Z_{ring,\,M}(\rho,\sigma;\rho_0) = 8\pi g_s M \alpha'^2
\,\left([b(\rho_0)]_+-b(\rho_0)\right) \,\frac{\sigma^2+ \rho^2
+\rho_0^2}{\left[(\sigma^2+ \rho^2 +\rho_0^2)^2-4\rho_0^2\rho^2
\right]^{3/2}}~,
\end{equation}
where $\rho=|\vec{y}|$ and $\sigma=|\vec{x}|$.  Fractional branes at
the enhan\c con ring, being tensionless, do not contribute directly to
the warp factor. Finally, there is a term sourced by the twisted field
strengths \be\label{twistedfluxes} d\gamma=\frac{i M}{\pi}
\frac{dz}{z} \,\Theta(|z|-|z_e|)\,\Theta(|z_0|-|z|)~.  
\ee In general
it takes the form
\begin{equation}\label{warp_integral}
Z_{fl}(\vec{y},\vec{x})=4\pi\alpha'^2 g_s^2 \int d^2
z\,|\partial_z\gamma|^2 \,\frac{1}{\left[
|\vec{x}|^2+|\vec{y}-\vec{z}|^2 \right]^2}~,
\end{equation}
which in the case under consideration reduces to
\begin{equation}\label{warp_fluxes_PRZ}
\begin{split}
Z_{fl,\, M}(\rho,r;\rho_e,\rho_0) &= \frac{2(g_sM\alpha')^2}{r^4}\,
\Bigg\{ 2\log
\frac{r^4+\left(\rho_e^2+\sqrt{(r^2+\rho_e^2)^2-4\rho_e^2\rho^2}\right)r^2-2\rho_e^2\rho^2
}{r^4+\left(\rho_0^2+\sqrt{(r^2+\rho_0^2)^2-4\rho_0^2\rho^2}\right)r^2-2\rho_0^2\rho^2
} +\\ &\hspace{-20pt}+2\log \frac{\rho_0^2}{\rho_e^2} +
\frac{r^2}{r^2-\rho^2} \left[
\frac{3(r^2-\rho^2)+\rho_0^2-\rho^2}{\sqrt{(r^2+\rho_0^2)^2-4\rho_0^2\rho^2}}-
\frac{3(r^2-\rho^2)+\rho_e^2-\rho^2}{\sqrt{(r^2+\rho_e^2)^2-4\rho_e^2\rho^2}}\right]\Bigg\}~,
\end{split}
\end{equation}
where $r^2=\rho^2+\sigma^2$.

Notice that the total D3 brane charge, which is $N+M$ if the UV theory
has gauge group $SU(N+M)\times SU(N+M)$, gets sectioned in different
pieces. The flux term carries a charge $M\,b(\rho_0)$, since
$b(\rho_0)=\frac{g_sM}{\pi}\,\log\frac{\rho_0}{\rho_e}$; the cutoff
anti-fractional branes carry a charge $M\left([b(\rho_0)]_+ -b(\rho_0)
\right)$; finally, there are $N-[b(\rho_0)]_-\,M$ regular D3 branes at
the origin.  This can be checked via the large $r$ asymptotics of the
different terms in the warp factor.

The vacuum considered in \cite{Petrini:2001fk} and described in
section \ref{sec:enhancon and SW} has $N$ regular D3 branes at the
origin, the enhan\c con ring at
$\rho_1=e^{-\frac{\pi}{2g_sM}}\,\rho_0$, and $M$ cutoff
anti-fractional branes at $\rho_0$, where $b(\rho_0)=\frac{1}{2}$,
carrying $M/2$ units of D3 charge; the twisted fluxes between
fractional and anti-fractional branes carry other $M/2$ units of D3
charge.

The vacuum with a finite cascade starting at $z_0$ and reaching
$SU(M)$ in the infrared has no regular D3 branes at the origin, $M$
fractional branes with no D3 charge melted at an enhan\c con ring at
$\rho_\mathrm{min} = \rho_{N/M}\equiv e^{-\frac{\pi
N}{g_sM^2}}\,\rho_1$, and $M$ cutoff anti-fractional branes at
$\rho_0$, where $b(\rho_0)=\frac{N}{M}+\frac{1}{2}$, carrying again
$M/2$ units of D3 charge; this time the twisted fluxes  between
fractional and anti-fractional branes carry  $N+M/2$ units of D3
charge. As we explained in detail in section \ref{sec:bearing}, what
happens is that at each generalized enhan\c con ring scale along the
cascade (where $b\in\mathbb{Z}$) melted tensionless fractional and
anti-fractional branes are left, naively annihilating if $c$ is
continuous crossing radially the generalized enhan\c con ring.  In
case $N=lM+p$ is not a multiple of $M$, then $\rho_\mathrm{min} =
e^{-\frac{\pi l}{g_sM}}\,\rho_1$, $b(\rho_0)=l +\frac{1}{2}$ and there
are $p$ D3 branes at the origin: the IR theory below the enhan\c con
scale is the $SU(p)\times SU(p)$ theory with one infinite coupling.

The infinite cascade limit can even be defined  continuously: it is
enough to send continuously the cutoff $\rho_0\to\infty$ keeping
$\rho_\mathrm{min}$ fixed and $b(\rho_\mathrm{min})=0$. This can be
achieved if $b(\rho_0)=\frac{g_s
M}{\pi}\ln\frac{\rho_0}{\rho_\mathrm{min}}$: as we change the cutoff
$\rho_0$, we also change the value of the gauge couplings at the
cutoff (and on the string side the tension of the cutoff branes) so
that low energy physics is not modified. Notice that every time a
$b(\rho_0)\in\bbZ$ threshold is crossed, the total D3 brane charge of
the configuration (the ranks of the UV CFT) jumps by $M$ units, and
the cutoff anti-fractional branes change.  The warp factor for the
infinite cascade with no regular D3 branes is nothing but $Z_{fl,\,
M}(\rho,r;\rho_\mathrm{min},\infty)$, see
eq. \eqref{warp_fluxes_PRZ}. If needed, the addition of $p$ regular D3
branes is straightforward.

We can also find the warp factor for a configuration with any number
of rotationally symmetric bearings. The total warp factor is sourced
by twisted fluxes and possibly by cutoff anti-fractional branes, if
there is no infinite cascade in the UV. Inside bearings fluxes vanish,
whereas outside they take the usual form
$|d\gamma|=\frac{M}{\pi}\,\frac{d\rho}{\rho}$. Therefore fluxes
contribute to the warp factor by a sum of terms taking the schematic
form $Z_{fl,\, M}(\rho,r;\rho^{(i+1)}_>,\rho^{(i)}_<)$, where
$\rho^{(i+1)}_>$ is the outer radius of the $(i+1)$-th bearing and
$\rho^{i}_<$ is the inner radius of the $i$-th bearing, if the
ordering points inwards. The requirement that $\rho^{(i)}_<$ and
$\rho^{(i+1)}_>$ be boundaries of subsequent bearings translates into
$\rho^{(i)}_<= e^\frac{\pi n_i}{g_s M}  \,\rho^{(i+1)}_>$, for some
$n_i\in\mathbb{N}$.

Finally, by now it should also be clear how to write the warp factor
in the case of perturbative Higgsings by backreacting rings of
tensionful fractional and anti-fractional branes, adding terms like
\eqref{warp_ring} sourced at suitable radii and with the suitable
normalizations.

\medskip

We end this section with some important remarks about the backreacted
geometries. For concreteness, we concentrate on solutions without
bearings nor perturbative Higgsing except at the cutoff, since the
generalization of the statements we are about to make should be clear.

\begin{figure}[tn]
\vskip -20pt
\centering \includegraphics[width=8.5cm]{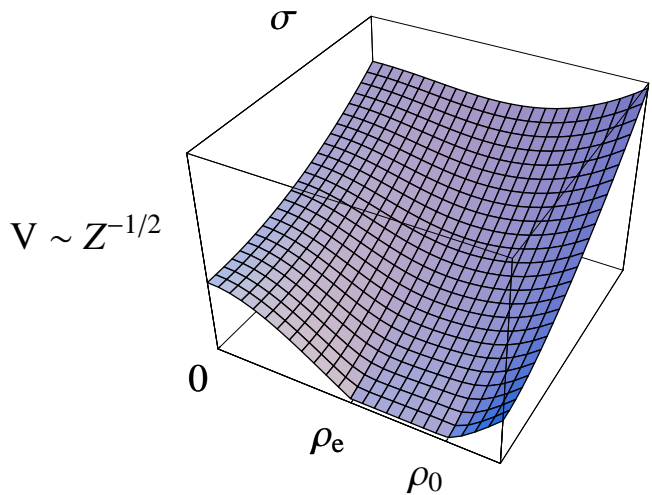}
\vskip -20pt
\caption{\small Potential $V=V(\rho,\sigma)$ felt by a massive particle
in the background dual to a vacuum with finite cascade. The origin is a saddle 
point while the absolute minimum is on the $\sigma=0$ axis all along the range where the dual field
theory undergoes a RG flow, from the enhan\c{c}on radius $\rho_e$ up to the UV cut-off $\rho_0$. 
\label{fig: potential}}
\end{figure}

The warp factor diverges (and the gravitational potential felt by a
massive particle has an
absolute minimum) only at the locations of sources for it (fractional
branes and twisted field strengths), namely on the orbifold plane
$\sigma=0$ and for $\rho\in[\rho_e,\rho_0]$. There are no repulsive
regions even when the D3 brane charge vanishes at some IR scale, as
occurs at the enhan\c con scale in the vacuum of \cite{Petrini:2001fk}
with $N=0$ and in the finite or infinite cascade solution with
$p=0$. Massive objects (but BPS ones) are always attracted by the sources 
of stress-energy: they want to go where twisted fluxes and fractional branes (and
possibly regular D3 branes) lie. For concreteness, we report in
figure \ref{fig: potential} the shape of the effective potential 
$V(\rho,\sigma)$ felt by a massive particle: it is proportional to 
$Z^{-1/2}$, once the kinetic terms are normalized to be 
$(\frac{d\rho}{d\tau})^2 + (\frac{d\sigma}{d\tau})^2$, $\tau$ 
being the worldline proper time. 

In these solutions the curvature diverges approaching the domain where twisted fluxes have support. Therefore, 
strictly speaking, the gravity solution cannot be trusted in that region and string theory is needed to resolve 
the curvature singularity. Still, the M-theory picture suggests that the form of the twisted fields will remain unchanged.

Finally, if there are no D3 branes
at the origin the geometry smoothly approaches flat space at $r=0$,
where the warp factor approaches
\begin{equation}\label{warp_origin_noD3}
Z(\mathbf{0})=2(g_sM\alpha')^2\,\left(
\frac{1}{\rho_e^4}-\frac{1}{\rho_0^4} \right) + 8\pi g_s M
\alpha'^2\left( [b(\rho_0)]_+-b(\rho_0)\right) \frac{1}{\rho_0^4}\;,
\end{equation}
signaling that excitations in the non-abelian sector have a minimal 
energy (consistently with the $SU(M)$ factor being broken to $U(1)^
{M-1}$). If instead there are regular D3 branes at the origin, they dominate 
the IR asymptotics which is $AdS_5\times S^5/\mathbb{Z}_2$, signaling a non-abelian fixed
point.


\section{Conclusions and outlook}
\label{sec: conc}

In this paper, we filled a gap in the understanding of the gauge theory dual interpretation of supergravity solutions 
with running fluxes, arising when considering fractional branes at generic Calabi-Yau singularities. 
It has been known for some time that fractional branes at isolated singularities describe RG flows which can be 
described in terms of cascades of Seiberg dualities. A similar interpretation was not possible for branes at 
non-isolated singularities, since their effective dynamics is intrinsically $\calN=2$. 

The basic outcome of our analysis is that, for branes at non-isolated singularities,
the reduction of the gauge group ranks along the RG flow can 
be understood in terms of a sequence of strong coupling transitions reminiscent of the low energy description
of the baryonic root of $\calN=2$ SQCD. The energy range spanned by the cascade
depends on the point in the Coulomb branch one is sitting at; specifically, on the number 
of non-vanishing VEV's for the adjoint scalars. 

We were also able to provide a gravity dual description of a new set of infinitely many vacua, characterized 
by new geometric structures, the enhan\c{c}on bearings, where the dual gauge theory alternates energy ranges where it runs, 
with ranges in which it is in a strongly coupled superconformal phase.

For all these vacua, an enhan\c{c}on mechanism takes place in the far IR. This changes the twisted fields configuration 
and ultimately the metric, whose correct repulson-free expression we provided for all vacua we have been studying. 

Our analysis focused, for definiteness, on the $A_1$ singularity, but our results have a much 
wider validity. First, they trivially extend to any $\calN=2$ singularity, as for instance the full 
ADE series. Second, any Calabi-Yau cone with non-isolated singularities, which upon the
inclusion of branes  generically gives rise to a  \Nugual{1} theory, should
present the same behavior. This is suggested from the supergravity solution and it is a rather 
non-trivial claim since SW techniques are not available in the $\calN=1$ context.

More complicated flows occur when fractional branes at isolated and non-isolated singularities 
are both present, which is in fact the most generic situation. Such a setup was recently considered 
in \cite{Argurio:2008mt} for a \Nugual{1} non-chiral $\bbZ_2$ orbifold of the conifold, where 
the complete RG flow of the dual cascading theory was extracted from supergravity. The analysis 
revealed that, while most of the rank reductions are understood in terms of Seiberg
duality, some of them cannot as the theory, due to the presence of adjoint fields, 
exhibits at some energies an effective \Nugual{2} behavior. We conjecture that in those cases too the rank reduction
is due to the adjoint fields being at baryonic-root-like points of its moduli space. One can easily follow 
on the field theory side all the cascades extracted from supergravity in \cite{Argurio:2008mt}, finding perfect agreement with our proposal.

It has been proposed in \cite{Argurio:2006ny,Argurio:2007qk} that gravity duals of metastable 
dynamical supersymmetry breaking models involve systems where fractional branes at isolated and 
non-isolated singularities are both present. The repulson-free warp factors we have found  
may prove useful to check those claims further, since the dynamics of $\calN=2$-like branes seems 
crucial to describe the metastable vacua. 

Finally, in view of our results, it would also be interesting to reconsider $\calN=2$ D3-D7 fractional 
brane systems and the corresponding gauge theory RG flows suggested by the known gravity duals \cite{Bertolini:2001qa}. 

\medskip

\medskip

\medskip

\centerline{\bf \large Acknowledgments}

\medskip
We would like to single out Riccardo Argurio for collaboration at the beginning of this project, continuous exchange 
of ideas and comments, and Alberto Zaffaroni for important remarks and insights. We are also grateful to 
Marco Bill\`o, Jarah Evslin, Igor Klebanov, Stanislav Kuperstein, 
Andrei Mikhailov, Yaron Oz, Rodolfo Russo, Cobi Sonnenschein and Shimon Yankielowicz for useful discussions. Finally, we are 
grateful to Ofer Aharony for suggestions and comments on a preliminary version of this paper.
F.B. acknowledges the support of the US Department of Energy under Grant No. DE-FG02-91ER40671.
C.C. is a Boursier FRIA-FNRS. The research of C.C. is also supported by IISN - Belgium (convention 4.4505.86)
and by the ``Interuniversity Attraction Poles Programme Belgian Science Policy".
The work of S.C. was supported in part by a center of excellence supported by the Israel Science Foundation
(grant No. 1468/06), by a grant (DIP H52) of the German 
Israel Project Cooperation, by a BSF United States-Israel binational science foundation grant
2006157 and German Israel Foundation (GIF) grant No. 962-94.7/2007.

\appendix

\section{Effective field theory approach to the cascading SW curve} \label{check double points}

Let us check the statements of section \ref{sec:SW curve cascading
vacuum} concerning the RG flow and the double points (\ref{dblpoints
1})-(\ref{dblpoints 2}), using an effective field theory approach for
the Seiberg-Witten curve between two strong coupling
transitions. Defining $\xi\equiv v^{M}$ and $\alpha \equiv
z_{0}^M$, we have the following Seiberg-Witten curve,
\be\label{SW curve in xi}
  \frac{\xi \prod_{j=0}^{h-1}
(\xi^2 + q^{\frac{1}{2}+2j} \alpha^2)}{(\xi-\alpha)\prod_{j=0}^{h-1}
(\xi^2 + q^{\frac{3}{2}+2j} \alpha^2)} = g(t|q) =
q^{\frac{1}{4}}(t+\frac{1}{t}) +\calO(q^{\frac{5}{4}})
\ee Now, defining $\Lambda_j^{2M} \equiv q^{j+\frac{1}{2}}z_{0}^{2M}$,
we can look at the curve in the range $\Lambda_{2n}<v<\Lambda_{2n-1}$, where, at small $q$,
\be
\frac{R}{S} \approx - \frac{q^{\frac{1}{4}} (\xi^2+\Lambda_{2n}^{2M})}{\Lambda_{2n}^M\xi} = g(t,q) \approx q^{\frac{1}{4}} (t+\frac{1}{t})\, ,
\ee
which gives
\be
\xi \Lambda_{2n}^M  t^2 +  (\xi^2+\Lambda^{2M}_{2n})t +\xi \Lambda_{2n}^M =0,
\ee
This is a SW curve for a $SU(2M)$ gauge group with $2M$ massless flavors
\cite{Witten:1997sc}, at the baryonic root (hence it has exact double points).
Extracting the roots for $t$ (and neglecting $\Lambda_n/v$ because of large $M$), one finds
\be\label{double values of u}
u_1 =-u_2= - \frac{M}{2\pi i} \log{\left( \frac{v}{\Lambda_{2n}} e^{-\frac{2\pi i k}{2M}} \right)} =
- \frac{M}{2\pi i} \log{\frac{v}{\Lambda_{2n}}} + \frac{1}{2}k,
\ee
where $k=0,1$. We see that at $v=\Lambda_{2n}$, $u_1=u_2=0,\frac{1}{2}$
(that is, the two NS5's intersect at $x^6=0$, but in fact the
corresponding M5 brane also self-intersects at two distinct points on
the torus). Since $\tau_1 = u_2-u_1= \tau-\tau_2$, we have reproduced the
correct perturbative running of the gauge couplings.
Also notice that this effective field theory for the first node is valid
only up to $v=\Lambda_{2n-1}$, where according to (\ref{double values
of u}) $u_1=u_2=\frac{\tau}{2}, \frac{\tau}{2}+\frac{1}{2}$ (that is
when the coupling of the second gauge group hits a Landau pole).

One can perform the same analysis for the second gauge group, i.e. for
the double points at $u=\frac{\tau}{2}, \frac{\tau}{2}+\frac{1}{2}$,
obtaining (\ref{dblpoints 2}).


\begin{thebibliography}{99}


\bibitem{Gubser:1998fp}
  S.~S.~Gubser and I.~R.~Klebanov,
  ``Baryons and domain walls in an $\calN = 1$ superconformal gauge theory,''
  Phys.\ Rev.\  D {\bf 58}, 125025 (1998)
  [arXiv:hep-th/9808075].

\bibitem{Klebanov:1999rd}
  I.~R.~Klebanov and N.~A.~Nekrasov,
  ``Gravity duals of fractional branes and logarithmic RG flow,''
  Nucl.\ Phys.\  B {\bf 574}, 263 (2000)
  [arXiv:hep-th/9911096].

\bibitem{Klebanov:2000nc}
  I.~R.~Klebanov and A.~A.~Tseytlin,
  ``Gravity duals of supersymmetric $SU(N) \times SU(N+M)$ gauge theories,''
  Nucl.\ Phys.\  B {\bf 578}, 123 (2000) [arXiv:hep-th/0002159].

\bibitem{Klebanov:2000hb}
  I.~R.~Klebanov and M.~J.~Strassler,
  ``Supergravity and a confining gauge theory: Duality cascades and
  $\chi$SB-resolution of naked singularities,'' JHEP {\bf 0008}, 052 (2000) [arXiv:hep-th/0007191].

\bibitem{Strassler:2005qs}
  M.~J.~Strassler,
  ``The duality cascade,''
  arXiv:hep-th/0505153.

\bibitem{Berenstein:2005xa}
  D.~Berenstein, C.~P.~Herzog, P.~Ouyang and S.~Pinansky,
  ``Supersymmetry breaking from a Calabi-Yau singularity,''
  JHEP {\bf 0509}, 084 (2005)
  [arXiv:hep-th/0505029].

\bibitem{Franco:2005zu}
  S.~Franco, A.~Hanany, F.~Saad and A.~M.~Uranga,
  ``Fractional branes and dynamical supersymmetry breaking,''
  JHEP {\bf 0601}, 011 (2006)
  [arXiv:hep-th/0505040].

\bibitem{Bertolini:2005di}
  M.~Bertolini, F.~Bigazzi and A.~L.~Cotrone,
  ``Supersymmetry breaking at the end of a cascade of Seiberg dualities,''
  Phys.\ Rev.\  D {\bf 72}, 061902 (2005)
  [arXiv:hep-th/0505055].

\bibitem{Intriligator:2005aw}
  K.~A.~Intriligator and N.~Seiberg,
  ``The runaway quiver,''
  JHEP {\bf 0602}, 031 (2006)
  [arXiv:hep-th/0512347].

\bibitem{Bertolini:2000dk}
  M.~Bertolini, P.~Di Vecchia, M.~Frau,
  A.~Lerda, R.~Marotta and I.~Pesando,
  ``Fractional D-branes and their gauge duals,''
  JHEP {\bf 0102}, 014 (2001)
  [arXiv:hep-th/0011077].

\bibitem{Polchinski:2000mx}
  J.~Polchinski, ``$\calN = 2$ gauge-gravity
  duals,'' Int.\ J.\ Mod.\ Phys.\  A {\bf 16}, 707 (2001)
  [arXiv:hep-th/0011193].

\bibitem{Aharony:2000pp}
  O.~Aharony, ``A note on the holographic
  interpretation of string theory backgrounds  with varying flux,''
  JHEP {\bf 0103}, 012 (2001) [arXiv:hep-th/0101013].

\bibitem{Petrini:2001fk}
  M.~Petrini, R.~Russo and A.~Zaffaroni,
  ``$\calN =  2$ gauge theories and systems with fractional branes,''
  Nucl.\ Phys.\ B {\bf 608}, 145 (2001) [arXiv:hep-th/0104026].

\bibitem{Benini:2007gx}
  F.~Benini, F.~Canoura, S.~Cremonesi, C.~Nunez
  and A.~V.~Ramallo, ``Backreacting Flavors in the Klebanov-Strassler
  Background,''
  JHEP {\bf 0709}, 109 (2007) [arXiv:0706.1238[hep-th]].

\bibitem{Argurio:2008mt}
  R.~Argurio, F.~Benini, M.~Bertolini,
  C.~Closset and S.~Cremonesi, ``Gauge/gravity duality and the
  interplay of various fractional branes,'' Phys.\ Rev.\  D {\bf 78},
  046008 (2008) [arXiv:0804.4470 [hep-th]].

\bibitem{Klebanov:1998hh}
  I.~R.~Klebanov and E.~Witten,
  ``Superconformal field theory on threebranes at a Calabi-Yau
  singularity,'' Nucl.\ Phys.\ B {\bf 536}, 199 (1998)
  [arXiv:hep-th/9807080].

\bibitem{Hollowood:2004ek}
  T.~J.~Hollowood and S.~Prem Kumar,
  JHEP {\bf 0412}, 034 (2004)
  [arXiv:hep-th/0407029].

\bibitem{Seiberg:1994aj}
  N.~Seiberg and E.~Witten, ``Monopoles,
  duality and chiral symmetry breaking in $\calN=2$ supersymmetric
  QCD,'' Nucl.\ Phys.\  B {\bf 431}, 484 (1994)
  [arXiv:hep-th/9408099].

\bibitem{Seiberg:1994rs}
  N.~Seiberg and E.~Witten, ``Electric -
  magnetic duality, monopole condensation, and confinement in
  $\calN=2$ supersymmetric Yang-Mills theory,''
  Nucl.\ Phys.\  B {\bf 426}, 19 (1994) [Erratum-ibid.\  B {\bf 430}, 485 (1994)]
  [arXiv:hep-th/9407087].

\bibitem{Argyres:1996eh}
  P.~C.~Argyres, M.~R.~Plesser and N.~Seiberg,
  ``The Moduli Space of $\calN=2$ SUSY {QCD} and Duality in $\calN=1$
  SUSY {QCD},'' Nucl.\ Phys.\  B {\bf 471}, 159 (1996)
  [arXiv:hep-th/9603042].

\bibitem{Johnson:1999qt}
  C.~V.~Johnson, A.~W.~Peet and J.~Polchinski,
  ``Gauge theory and the excision of repulson singularities,''
  Phys.\ Rev.\  D {\bf 61}, 086001 (2000) [arXiv:hep-th/9911161].

\bibitem{Witten:1997sc}
  E.~Witten, ``Solutions of four-dimensional
  field theories via M-theory,'' Nucl.\ Phys.\  B {\bf 500}, 3 (1997)
  [arXiv:hep-th/9703166].

\bibitem{Kachru:1998ys}
  S.~Kachru and E.~Silverstein,
  ``4d conformal theories and strings on orbifolds,''
  Phys.\ Rev.\ Lett.\  {\bf 80}, 4855 (1998)
  [arXiv:hep-th/9802183].

\bibitem{Diaconescu:1997br}
  D.~E.~Diaconescu, M.~R.~Douglas and
  J.~Gomis, ``Fractional branes and wrapped branes,'' JHEP {\bf 9802},
  013 (1998) [arXiv:hep-th/9712230].

\bibitem{Grana:2001xn}
  M.~Grana and J.~Polchinski, ``Gauge / gravity
  duals with holomorphic dilaton,'' Phys.\ Rev.\  D {\bf 65}, 126005
  (2002) [arXiv:hep-th/0106014].

\bibitem{Douglas:1995nw}
  M.~R.~Douglas and S.~H.~Shenker, ``Dynamics
  of $SU(N)$ supersymmetric gauge theory,''
  Nucl.\ Phys.\  B {\bf 447}, 271 (1995) [arXiv:hep-th/9503163].

\bibitem{Witten:1995zh}
  E.~Witten, ``Some comments on string
  dynamics,'' arXiv:hep-th/9507121.

\bibitem{Aspinwall:1996mn}
  P.~S.~Aspinwall, ``K3 surfaces and string
  duality,'' arXiv:hep-th/9611137.

\bibitem{Blum:1997fw}
  J.~D.~Blum and K.~A.~Intriligator,
  ``Consistency conditions for branes at orbifold singularities,''
  Nucl.\ Phys.\  B {\bf 506}, 223 (1997) [arXiv:hep-th/9705030].

\bibitem{Bertolini:2001qa}
  M.~Bertolini, P.~Di Vecchia, M.~Frau, A.~Lerda and R.~Marotta,
  ``N = 2 gauge theories on systems of fractional D3/D7 branes,''
  Nucl.\ Phys.\  B {\bf 621} (2002) 157
  [arXiv:hep-th/0107057].

\bibitem{Billo:2001vg}
  M.~Billo, L.~Gallot and A.~Liccardo,
  ``Classical geometry and gauge duals for fractional branes on ALE
  orbifolds,'' Nucl.\ Phys.\  B {\bf 614}, 254 (2001)
  [arXiv:hep-th/0105258].

\bibitem{Klemm:1994qs}
  A.~Klemm, W.~Lerche, S.~Yankielowicz and S.~Theisen,
  ``Simple singularities and $\calN=2$ supersymmetric Yang-Mills theory,''
  Phys.\ Lett.\  B {\bf 344}, 169 (1995) [arXiv:hep-th/9411048].

\bibitem{Argyres:1994xh}
  P.~C.~Argyres and A.~E.~Faraggi,
  Phys.\ Rev.\ Lett.\  {\bf 74}, 3931 (1995)
  [arXiv:hep-th/9411057].

\bibitem{Giveon:1998sr}
  A.~Giveon and D.~Kutasov,
  ``Brane dynamics and gauge theory,''
  Rev.\ Mod.\ Phys.\  {\bf 71}, 983 (1999)
  [arXiv:hep-th/9802067].

\bibitem{Ennes:1999fb}
  I.~P.~Ennes, C.~Lozano, S.~G.~Naculich and H.~J.~Schnitzer,
  ``Elliptic models and M-theory,''
  Nucl.\ Phys.\  B {\bf 576} (2000) 313
  [arXiv:hep-th/9912133].

\bibitem{Bolognesi:2008sw}
 S.~Bolognesi,
  ``A Coincidence Problem: How to Flow from N=2 SQCD to N=1 SQCD,''
  arXiv:0807.2456 [hep-th].

\bibitem{Hanany:1995na}
  A.~Hanany and Y.~Oz,
  ``On the quantum moduli
  space of vacua of $\calN=2$ supersymmetric $SU(N_c)$ gauge
  theories,'' Nucl.\ Phys.\  B {\bf 452}, 283 (1995)
  [arXiv:hep-th/9505075].
  
\bibitem{Argyres:1995wt}
  P.~C.~Argyres, M.~R.~Plesser and A.~D.~Shapere,
  ``The Coulomb phase of $\calN=2$ supersymmetric QCD,''
  Phys.\ Rev.\ Lett.\  {\bf 75}, 1699 (1995)
  [arXiv:hep-th/9505100].

\bibitem{Evslin:2004vs}
  J.~Evslin, ``The cascade is a MMS instanton,''
  arXiv:hep-th/0405210.

\bibitem{Hori:1997ab}
  K.~Hori, H.~Ooguri and Y.~Oz, ``Strong coupling
  dynamics of four-dimensional $\calN = 1$ gauge theories from
  M-theory fivebrane,'' Adv.\ Theor.\ Math.\ Phys.\  {\bf 1}, 1 (1998)
  [arXiv:hep-th/9706082].

\bibitem{Witten:1997ep}
  E.~Witten, ``Branes and the dynamics of
  {QCD},'' Nucl.\ Phys.\  B {\bf 507}, 658 (1997)
  [arXiv:hep-th/9706109].

\bibitem{Aharony:2005zr}
  O.~Aharony, A.~Buchel and A.~Yarom,
  ``Holographic renormalization of cascading gauge theories,''
  Phys.\ Rev.\  D {\bf 72}, 066003 (2005)
  [arXiv:hep-th/0506002].

\bibitem{deBoer:1997zy}
  J.~de Boer, K.~Hori, H.~Ooguri and Y.~Oz,
  ``Kaehler potential and higher derivative terms from M theory five-brane,''
  Nucl.\ Phys.\  B {\bf 518}, 173 (1998)
  [arXiv:hep-th/9711143].

\bibitem{Halmagyi:2004jy}
  N.~Halmagyi, K.~Pilch, C.~Romelsberger and
  N.~P.~Warner, ``The complex geometry of holographic flows of quiver
  gauge theories,'' JHEP {\bf 0609}, 063 (2006)
  [arXiv:hep-th/0406147].

\bibitem{Johnson:2001wm}
  C.~V.~Johnson, R.~C.~Myers, A.~W.~Peet and S.~F.~Ross,
  ``The enhan\c con and the consistency of excision,''
  Phys.\ Rev.\  D {\bf 64}, 106001 (2001)
  [arXiv:hep-th/0105077].

\bibitem{Fayyazuddin:1997by}
  A.~Fayyazuddin and M.~Spalinski,
  ``The Seiberg-Witten differential from M-theory,''
  Nucl.\ Phys.\  B {\bf 508}, 219 (1997)
  [arXiv:hep-th/9706087].

\bibitem{Henningson:1997hy}
  M.~Henningson and P.~Yi,
  ``Four-dimensional BPS-spectra via M-theory,''
  Phys.\ Rev.\  D {\bf 57}, 1291 (1998)
  [arXiv:hep-th/9707251].

\bibitem{Mikhailov:1997jv}
  A.~Mikhailov,
  ``BPS states and minimal surfaces,''
  Nucl.\ Phys.\  B {\bf 533}, 243 (1998)
  [arXiv:hep-th/9708068].

\bibitem{Argurio:2006ny}
  R.~Argurio, M.~Bertolini, S.~Franco and
  S.~Kachru, ``Gauge/gravity duality and meta-stable dynamical
  supersymmetry breaking,'' JHEP {\bf 0701}, 083 (2007)
  [arXiv:hep-th/0610212].

\bibitem{Argurio:2007qk}
  R.~Argurio, M.~Bertolini, S.~Franco and
  S.~Kachru, ``Metastable vacua and D-branes at the conifold,'' JHEP
  {\bf 0706}, 017 (2007) [arXiv:hep-th/0703236].





\end{thebibliography}
\end{document}